\documentclass[fleqn,usenatbib,useAMS]{mnras}
\usepackage{graphicx}	
\usepackage{amsmath}	
\usepackage{amssymb}	
\usepackage{multicol}        
\usepackage{bm}		

\usepackage{color}
\usepackage{pdflscape}
\usepackage[T1]{fontenc}
\usepackage{ae,aecompl}
\usepackage{newtxtext,newtxmath}
\usepackage{lscape}
\usepackage{mathtext}
\usepackage{float}
\usepackage{textcomp}
\usepackage{rotating}
\usepackage{dcolumn}
\usepackage{tabularx}
\usepackage{pdflscape} 
\usepackage{url}
\usepackage{textcomp}

\renewcommand{\theenumi}{\roman{enumi}}%

\def\aj{Astron. J.}
\def\apj{Astrophys. J.}
\def\apjl{Astrophys. J. Lett.}
\def\apjs{Astrophys. J. Suppl. Ser.}

\def\aap{Astron Astrophys.}

\def\araa{ARAA}
\def\mnras{MNRAS}
\def\nat{Nature}

\def\sovast{Sov. Astron.}

\newcommand{\Msun}{M_\odot}
\newcommand{\Mbh}{M_{\rm BH}}

\newcommand{\ledd}{\lambda_{\rm Edd}}

\newcommand{\lx}{\lambda_{\rm X}}
\newcommand{\lxs}{\langle \lambda_{\rm X} \rangle}
\newcommand{\Lx}{L_{\rm X}}
\newcommand{\Lxs}{\langle L_{\rm X} \rangle}
\newcommand{\Fx}{F_{\rm X}}
\newcommand{\sigmax}{\sigma_{\rm X}}
\newcommand{\Fmeanx}{\langle \Fx \rangle}

\newcommand{\Le}{L_{\rm Edd}}
\newcommand{\Lbol}{L_{\rm bol}}

\newcommand{\wi}{\mathrm{w}_i}
\newcommand{\whi}{\hat{\mathrm{w}}_i}

\newcommand{\nh}{N_{\rm H}}
\newcommand{\nhg}{N_{\rm H, Gal}}
\newcommand{\kx}{k_{\rm X}}
\newcommand{\Fr}{F_{\rm r}}
\newcommand{\Ftrue}{F_{\rm X,true}}
\newcommand{\ftrue}{f_{\rm true}}
\newcommand{\tobs}{t_{\rm obs}}
\newcommand{\dtobs}{\Delta t_{\rm obs}}
\newcommand{\npair}{N_{\rm pairs}}

\newcommand{\torb}{t_{\rm orb}}
\newcommand{\tth}{t_{\rm th}}
\newcommand{\tvisc}{t_{\rm visc}}
\newcommand{\Rs}{R_{S}}

\newcommand{\nub}{\nu_{\rm b}}

\newcommand{\srg}{{\it SRG}}
\newcommand{\xmm}{{\it XMM-Newton}}
\newcommand{\rosat}{{\it ROSAT}}
\newcommand{\chandra}{{\it Chandra}}

\title[Quasar X-ray variability]{X-ray variability of SDSS quasars based on the SRG/eROSITA all-sky survey}

 \author[Prokhorenko~et~al.]{Prokhorenko S.A.$^{1,2}$\thanks{E-mail: \href{mailto:sprokhorenko@cosmos.ru}{sprokhorenko@cosmos.ru}}, 
 Sazonov S.Yu.$^{1,2}$,
 Gilfanov M.R.$^{1,3}$, 
 Balashev S.A.$^{4}$, 
 Bikmaev I.F.$^{5,6}$, 
 Ivanchik A.V.$^{4}$, 
 \newauthor
 Medvedev P.S.$^{1}$,
 Starobinsky A.A.$^{7,2}$, 
 Sunyaev R.A.$^{1,3,8}$\\
$^{1}$ Space Research Institute of the Russian Academy of Sciences, Profsoyuznaya Str. 84/32, 117997 Moscow, Russia\\
$^{2}$ National Research University Higher School of Economics, Pokrovsky Bulvar 11, 101990 Moscow, Russia\\
$^{3}$ Max-Planck-Institut f\"{u}r Astrophysik, Karl-Schwarzschild-Str. 1, D-85741 Garching, Germany \\
$^{4}$ Ioffe Institute, Politeknicheskaya str. 26, St Petersburg 194021, Russia\\
$^{5}$ Kazan Federal University, Kremlevskaya str. 18, 420008 Kazan, Russia\\
$^{6}$ Academy of Sciences of Tatarstan, Baumana Str. 20, 420111 Kazan, Russia\\
$^{7}$ L.D. Landau Institute for Theoretical Physics of the Russian Academy of Sciences, Chernogolovka, 142432 Moscow region, Russia\\
$^{8}$ Institute for Advanced Study, 1st Einstein Drive, Princeton, NJ, 08540, USA.
}
\date{Accepted XXX. Received YYY; in original form ZZZ}

\pubyear{2023}

\begin{document}

\newcommand{\prange}[3][2]{10^{#2} < #1 < 10^{#3}}

\label{firstpage}

\maketitle

\begin{abstract}
We examine the long-term (rest-frame time scales from a few months to $\sim 20$\,years) X-ray variability of a sample of 2344 X-ray bright quasars from the SDSS DR14Q Catalogue, based on the data of the \srg/eROSITA All-Sky Survey complemented for $\sim 7$\% of the sample by archival data from the \xmm\ Serendipitous Source Catalogue. We characterise variability by a structure function, $SF^2(\Delta t)$. We confirm the previously known anti-correlation of the X-ray variability amplitude with luminosity. We also study the dependence of X-ray variability on black hole mass, $\Mbh$, and on an X-ray based proxy of the Eddington ratio, $\lx$. Less massive black holes prove to be more variable for given Eddington ratio and time scale. X-ray variability also grows with decreasing Eddington ratio and becomes particularly strong at $\lx$ of less than a few per cent. We confirm that the X-ray variability amplitude increases with increasing time scale. The $SF^2(\Delta t)$ dependence can be satisfactorily described by a power law, with the slope ranging from $\sim 0$ to $\sim 0.4$ for different ($\Mbh$, $\lx$) subsamples (except for the subsample with  the lowest black hole mass and lowest Eddington ratio, where it is equal to $1.1\pm 0.4$).
\end{abstract}
\section{Introduction}
\label{s:intro}

The X-ray emission due to accretion of matter onto a black hole (BH) is expected to be variable (e.g. \citealt{sunyaev1973,lightman1974,shakura1976,pringle1981}). By studying the variability, together with the spectral properties and, since recently, polarization \citep{ixpe2022} of the X-ray emission, one can get insight into the physics of the accretion disc and its hot corona and obtain constraints on the BH properties such as mass, spin and accretion rate.

Observations indeed reveal that virtually all BH stellar X-ray binaries and active galactic nuclei (AGN) are variable X-ray sources. Thus far, X-ray variability has been studied in greater detail for X-ray binaries (see \citealt{mcclintock2006,gilfanov2010,belloni2014} for reviews) than for AGN. This is partly due to the fact that BH X-ray binaries are among the brightest X-ray sources in the sky (their fluxes typically being $\sim 0.1$\,Crab), whereas even the brightest, nearby Seyfert galaxies usually have fluxes of a few mCrab, while quasars are usually yet fainter, which, of course, narrows the possibilities of their investigation. More important is that all characteristic time scales of accretion discs are 4-9 orders of magnitude longer for AGN than for X-ray binaries due to the huge difference in BH mass ($\sim 10^{5}$--$10^{10}\Msun$ vs. $\sim 10\Msun$). In particular, the orbital, thermal and viscous times scale with the BH mass $\Mbh$ and radius $R$ in the disc as follows \citep{shakura1973}:
\begin{equation}
\torb\approx 3\left(\frac{\Mbh}{10^8\,\Msun}\right)\left(\frac{R}{10\,\Rs}\right)^{3/2}\,{\rm days},
\label{eq:torb}
\end{equation}
\begin{equation}
\tth=\frac{\torb}{2\pi\alpha}\approx 50\left(\frac{\alpha}{0.01}\right)^{-1}\left(\frac{\Mbh}{10^8\,\Msun}\right)\left(\frac{R}{10\,\Rs}\right)^{3/2}\,{\rm days}, 
\label{eq:tth}
\end{equation}
\begin{equation}
\tvisc=\left(\frac{H}{R}\right)^{-2}\tth,
\label{eq:tvisc}
\end{equation}
where $\Rs=2G\Mbh/c^2$ is the Schwarzschild radius of the BH, $G$ is the gravitational constant, $c$ is the speed of light, $\alpha$ is the viscosity parameter and $H$ is the scale height of the disc. Therefore, in AGN, it is usually possible to study variability on the dynamical and thermal time scales, but the viscous time scale (in the inner region of the accretion disc) can be probed for relatively light SMBHs only, given the limited duration of monitoring campaigns.  

X-ray variability of AGN is interesting not only from the point of view of studying the physics of accretion onto supermassive black holes (SMBHs), but also in the cosmological context due to the existence of correlation between the X-ray and UV luminosities of AGN \citep{tananbaum1979}. If this relation is non-linear, as many studies suggest (e.g. \citealt{strateva2005,lusso2016tight,salvestrini2019}), then there is an exciting possibility of using quasars as ``standard'' candles for determination of cosmological parameters \citep{risaliti2015hubble,risaliti2019}. However, there is a serious obstacle on this route, namely that the UV--X-ray correlation is characterised by a significant scatter. It is thus necessary to understand the origin(s) of this scatter and take it into account when applying this relation to cosmological measurements. Variability of quasars in X-rays and in the optical--UV bands is definitely one of the major causes of the observed scatter \citep{ulrich1997,vandenberk2004,kelly2009,vagnetti2013,caplar2017,chiaraluce2018,burke2021,arevalo2023}.

X-ray variability of AGN can be studied in different ways. One approach consists of conducting extensive observational campaigns of individual objects aimed at obtaining their detailed X-ray light curves and estimating their power spectral density (PSD), covering time scales from minutes to years. This approach is quite demanding in terms of observational time and sensitivity, and has thus far been implemented mostly for bright Seyfert galaxies (e.g. \citealt{mushotzky1993,nandra1997,uttley2002,markowitz2004,gonzalez2012}). These studies have shown that AGN PSDs can usually be described by a bending power law, with a slope of $\sim 2$ at high frequencies and a flatter slope of $\sim 0$--1 at lower frequencies, with the characteristic time scale varying between a few minutes and a few years from one Seyfert galaxy to another \citep{uttley2005a,vaughan2005,mchardy2006}. When compared to the X-ray variability properties of BH X-ray binaries, these results support the idea that AGN are scaled-up analogs of the latter.

In studying the X-ray variability of high-luminosity, distant AGN (quasars), one usually encounters the problem that the sampling and quality of the X-ray light curves is insufficient for constructing power-density spectra on a source-by-source basis. Nevertheless, given a large enough sample of objects, one can use another approach that consists of estimating some integral characteristics of X-ray variability, e.g. normalized excess variance, for each object and examining how these quantities depend upon quasar properties such as redshift, X-ray luminosity or BH mass as well as on the rest-frame time scale. Within this approach, it is often possible and desirable to bin the objects and/or measurements according to these physical properties, i.e. employ an ensemble-averaged approach, based on the implicit assumption that all AGN are intrinsically similar objects. 

This method has been successfully implemented by a number of authors, using quasar samples selected from deep extragalactic surveys by such X-ray observatories as \rosat, \xmm\ and \chandra\ \citep{almaini2000,papadakis2008,yang2016,paolillo2017,paolillo2023}. These studies indicated that the X-ray variability properties of quasars indeed depend on their physical characteristics such as BH mass and accretion rate (see a discussion in Section~\ref{s:discuss} below). However, the available statistics was usually not sufficient for constraining the dependencies on these parameters simultaneously, so that there is a need to continue this research using larger samples, with a better coverage of temporal frequencies, redshifts, BH masses and accretion rates.

The all-sky X-ray survey that has been conducted since December 2019 by the eROSITA telescope \citep{predehl2021} on board the \srg\ orbital observatory \citep{sunyaev2021} for the first time allows us to study the X-ray variability of quasars on a massive basis (using thousands of objects). In this paper, we report the results of such a variability analysis for a sample of X-ray bright quasars detected during the \srg/eROSITA all-sky survey and selected from the SDSS catalogue of optical quasars \citep{paris2018}. 

\section{SRG/eROSITA--SDSS sample of X-ray bright quasars}
\label{s:data}

We use a sample of X-ray bright quasars found by cross-correlating the catalogue of X-ray sources detected by \srg/eROSITA during the all-sky survey in the $0^\circ<l<180^\circ$ celestial hemisphere with the optical SDSS DR14 quasar catalogue (SDSS DR14Q, \citealt{paris2018}). The latter covers 9376 square degrees on the sky, 8548 of which are at $0^\circ<l<180^\circ$. 

Specifically, the X-ray sources were selected from the catalogue of point X-ray sources detected on the summed map of five eROSITA all-sky surveys\footnote{The fifth survey was interrupted on February 26th, 2022, so that the data of only four surveys are available for $\sim 60$\% of the sky.} in the 0.3--2.3\,keV energy band and were required to have an average (i.e. determined from the summed map) observed flux of at least $2 \times 10^{-13}$\,erg\,s$^{-1}$\,cm$^{-2}$ in this band. Our analysis of X-ray variability is based on the source fluxes measured in individual eROSITA sky surveys. These fluxes were evaluated by forced X-ray photometry, using the source positions determined from the summed eROSITA map (see \citealt{medvedev2022}).

The observed fluxes provided in the eROSITA catalogue are estimated from the measured count rates assuming a universal absorbed power-law spectral model with a photon index of $\Gamma=2.0$ and a Galactic column density of $\nh=3\times 10^{20}$\,cm$^{-2}$. We applied an approximate absorption correction to the observed fluxes adopting this column density, i.e. multiplied the fluxes by 1.18.  Hereafter, we denote the unabsorbed fluxes (in the observed 0.3--2.3\,keV energy band) in individual eROSITA sky surveys by $\Fx$ and the average eROSITA fluxes by $\Fmeanx$. The actual Galactic absorption, $\nhg$, in the direction of the different quasars varies significantly, with 90\% of the objects having $\nhg$ between $10^{20}$ and $5\times 10^{20}$\,cm$^{-2}$ (the median column density is $2\times 10^{20}$\,cm$^{-2}$). This corresponds to $\pm 10$\% variations in the flux absorption correction. We have not taken this minor factor into account in view of other possible, unaccounted for, uncertainties of the same order: namely, there may be source-to-source variations in the intrinsic slope of the X-ray continuum and additional absorption intrinsic to the quasars. This is justifiable, since a systematic uncertainty of 10\% in flux (and hence luminosity) is much smaller than the width of the luminosity bins ($\sim 1$\,dex) that we use below to study the dependence of X-ray variability on quasar luminosity. We also emphasise that absorption correction has no impact on the determination of the variability amplitude of each given source.

The SDSS DR14Q catalogue comprises spectroscopically studied objects that have been confirmed as quasars via an automated procedure combined with a partial visual inspection of spectra, have luminosities $M_i[z=2]< -20.5$, and either display at least one emission line with a full width at half maximum larger than 500\,km\,s$^{-1}$ or have interesting/complex absorption features. We conducted the search within the eROSITA 98\% positional uncertainty, defined by radius $r98$. For X-ray bright sources with $\Fmeanx>2\times 10^{-13}$\,erg\,s$^{-1}$\,cm$^{-2}$, this radius is typically $\sim 5\arcsec$ and does not exceed $6.5\arcsec$ for the \srg-SDSS X-ray bright quasar sample. We excluded a number of unreliable associations. First, we removed 10 objects that are either not present in the SDSS DR16Qprop catalogue \citep{wu2022catalog} or have redshift estimates in that catalogue that significantly differ from those in SDSS DR14Q\footnote{Although SDSS DR16Qprop is an extended version of SDSS DR14Q, we base our study on the latter since it is known to contain a smaller fraction of false redshifts due to the more stringent verification.}. Second, we excluded two pairs of quasars (four objects) that are separated from each other by less than 20\arcsec. The expected number of spurious eROSITA--SDSS matches is $\sim 1$ (it is less than 2 with a probability of 90\%), which is tolerable for our purposes. We obtained this estimate by assigning small (but much larger than $r98$) angular offsets to all SDSS DR14Q quasars and then counting the number of cross-matches with our eROSITA sample. 

To minimise the impact of exceptionally large (and thus significantly non-Gaussian) flux uncertainties on our variability analysis, we excluded measurements (151 in total) with the vignetting corrected eROSITA exposure time of less than 70 seconds (while the typical exposure time of a single observation is $\sim 200$ seconds). In addition, we excluded 11 $\Fx$ estimates for which the forced photometry yielded zero values. Such cases occasionally arise because the likelihood function that is used in the forced photometry is not defined for negative fluxes, and although this procedure also provides flux upper limits their consideration would unnecessarily complicate our X-ray variability analysis. Together, these two steps have removed just 162 (i.e. less than 2\%) of all flux measurements, which cannot affect the results of this study in any significant way.

Blazars (BL Lac objects and flat-spectrum radio quasars) are a special class of AGN, which, due to the presence of a highly collimated emission component, can have substantially different variability properties from normal (i.e. unbeamed) AGN. We thus tried to carefully clean our sample from blazars. First, we cross-correlated our sample with the 5th edition of the Roma-BZCAT Multifrequency Catalogue of Blazars \citep{massaro20155th}. This yielded 184 counterparts. Second, we cross-correlated our sample with the fourth catalogue of AGN detected by the Fermi Large Area Telescope, Data Release 3 \citep{ajello2022fourth}, namely with objects classified as `bll' (BL Lac-type objects), `bcu' (blazar candidates of unknown types) or `fsqr' (Flat Spectrum Radio Quasars). This resulted in an additional 10 counterparts. We then cross-matched our sample with the Blazar Radio and Optical Survey (BROS) \citep{itoh2020blazar}, excluding Gigahertz peaked-spectrum sources and compact steep-spectrum radio sources (based on the corresponding flag in the catalogue) from the search. This provided an additional 67 sources. Finally, we cross-correlated our sample with the Combined Radio All-Sky Targeted Eight GHz Survey \citep{healey2007crates}, which yielded another 18 counterparts. In all these cases, the search was done within 10\arcsec\ of the SDSS DR14Q optical positions of our objects to take into account that some of the positions given in the blazar catalogues are not very precise (in particular, when only a radio position is available). The expected total number of spurious cross-matches with the blazar catalogues is less than one.

In total, we found 279 blazars or blazar candidates associated with our sample of X-ray bright quasars (see Table~\ref{blazars} for details) and removed them from the sample. It is possible that some of the discarded objects are actually not blazars, but that does not present a problem since our preference is to achieve maximum purity of the quasar sample. 

The resulting clean sample consists of 2344 quasars. Of these, 1224 have five eROSITA flux measurements, 1074 have four, 45 have three and one object has only two fluxes.

\begin{table} 
	\vspace{2mm}
	\centering
	
	\vspace{2mm}
	\begin{tabular}{ m{5.9cm} | m{1.5cm} } \hline\hline
		{Catalogue}&  Number of counterparts \\ \hline
		{Roma-BZCAT Multifrequency Catalogue of Blazars}   &   184   \\
		{4th Fermi/LAT catalogue of AGN DR3}  &   74  \\
		{Blazar Radio and Optical Survey}      &    188    \\
		{Combined Radio All-Sky Targeted Eight GHz Survey} &    192\\
		{All of the above}      &    279  \\
		
        \hline
	\end{tabular}
    \caption{Cross-correlation with blazar catalogues.}
    \label{blazars}
\end{table}

Our choice of the relatively high X-ray flux limit ($2 \times 10^{-13}$\,erg\,s$^{-1}$\,cm$^{-2}$ before absorption correction) for this study is primarily driven by the desire to better control the impact of flux measurement uncertainties on the variability analysis. With this threshold, 93.4\% of the individual flux measurements in our sample are based on more than 20 net source counts, so that the corresponding flux uncertainties are expected to be close to Gaussian.\footnote{As described in \cite{medvedev2022}, eROSITA fluxes are obtained via maximum likelihood (ML) point-spread-function (PSF) fitting of counts images. The flux values for individual sky surveys used in the variability analysis are computed via ML fitting with a fixed source position (which is determined on the summed data of all surveys). Our simulations show that the probability distribution for ML flux is different from Poisson and becomes close to Gaussian at the net source counts exceeding $\sim 20$, similar to Poisson statistics. This is a consequence of the fact that ML flux for a fixed source position is determined by a linear combination of the pixel values with the weights determined by the PSF and the background level with respect to the source intensity.}

\begin{figure}
	\vspace{6mm}
	\includegraphics[width=\columnwidth]{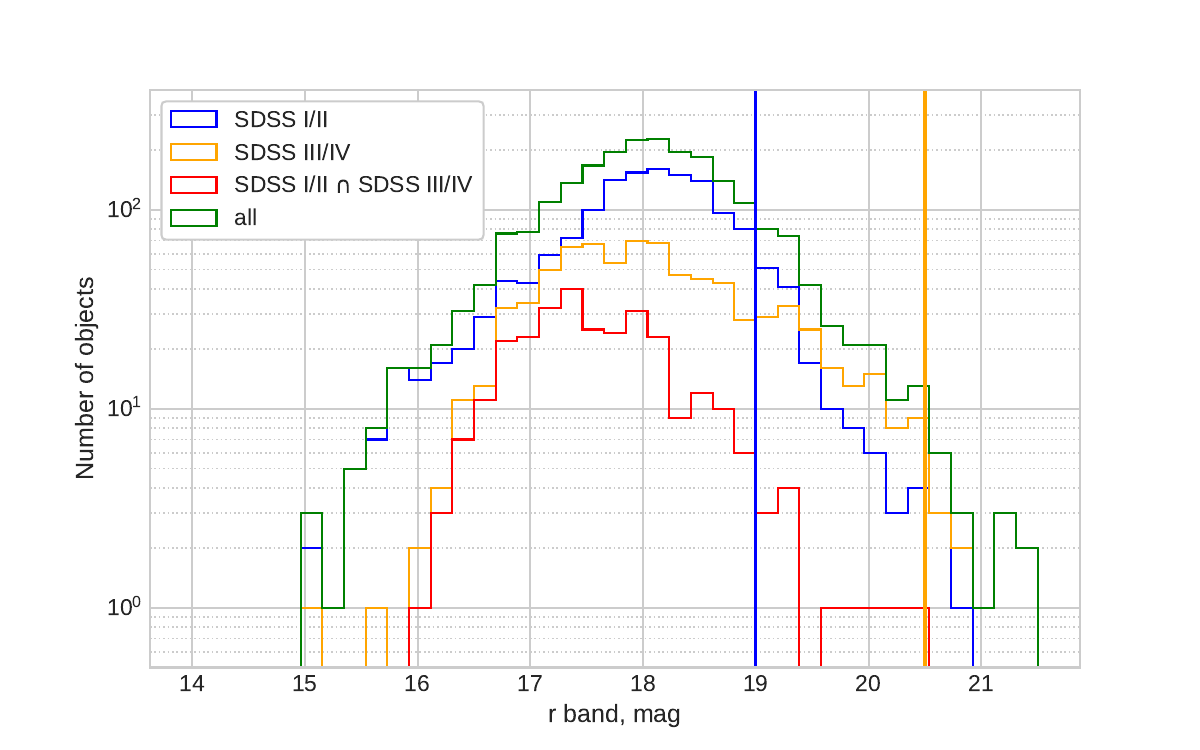}
	\caption{Distribution of the SDSS r-band apparent magnitudes, corrected for the Galactic extinction, of the \srg-SDSS sample of X-ray bright quasars studied in this work. The whole sample is shown in green. The other histograms show the following subsamples: blue, quasars observed in SDSS I/II only; orange, quasars observed in SDSS III/IV only; red, quasars observed both in SDSS I/II and SDSS III/IV. The vertical blue and orange lines indicate the effective spectroscopic depths of SDSS I/II and  SDSS III/IV. }
\label{rbanddistr} 
\end{figure}

Secondly, at this X-ray flux limit, our sample is characterised by high optical completeness. Figure~\ref{rbanddistr} shows the distribution of the apparent magnitudes of our X-ray selected quasars in the SDSS r band. The maximum of the distribution lies at $r\sim 18$, while the effective threshold of the SDSS DR14Q catalogue is significantly deeper. Namely, SDSS DR14Q is based on spectroscopy carried out during the SDSS I/II and SDSS III/IV surveys, which had different quasar selection criteria. Based on fig.~6 in \cite{paris2018}, we can estimate the effective completeness limit of SDSS I/II at $r\sim 19$ and that of SDSS III/IV at $r\sim 20.5$ (see the corresponding vertical lines in Fig.~\ref{rbanddistr}). Therefore, it is unlikely that we are missing a substantial fraction of X-ray bright quasars in the SDSS DR14Q footprint due to their faintness in the optical. Estimating this incompleteness in quantitative terms goes beyond the scope of this study and is not necessary here, since we are not exploring space densities of quasars. 

\begin{figure*}
    \begin{center}
    	\vspace{6mm}
    	\includegraphics[width=17cm]{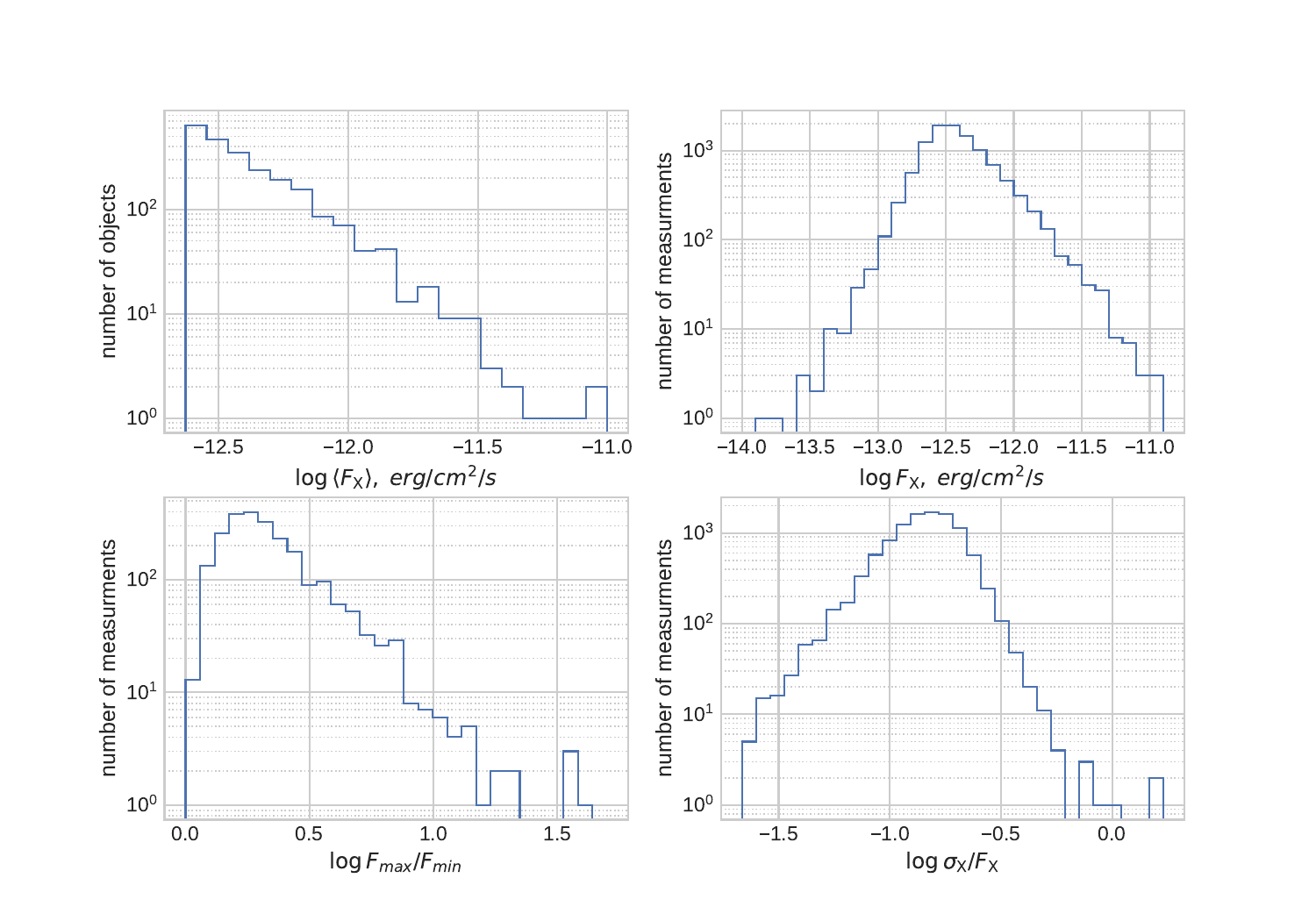}
    	\caption{Differential distributions of various X-ray properties of the \srg/eROSITA--SDSS X-ray bright quasar sample, based on eROSITA data. Upper left panel: time-averaged X-ray fluxes. Upper right panel: individual (per \srg\ sky survey) X-ray flux measurements. Lower left panel: ratio of the maximum to minimum X-ray fluxes. Lower right panel: relative uncertainties of individual flux measurements. Hereafter, $\log$ means $\log_{10}$.}
    \end{center}
    \label{Fdistrs}
\end{figure*}

Figure~\ref{Fdistrs} illustrates the main X-ray properties of our quasar sample. Namely, we show the distributions of: (i) average X-ray fluxes, (ii) individual flux measurements, (iii) ratios of the maximum to minimum flux, and (iv) relative flux uncertainties, all based on eROSITA data. For most of the objects, the X-ray flux varies by less than {\rm a factor of two} between the \srg\ sky surveys but some quasars demonstrate stronger variability. In particular, the flux varies more than tenfold for 27 sources. The relative flux uncertainty ($\sigmax/\Fx$) usually does not exceed 20\%. Hereafter $\sigmax$ denotes the $1\sigma$ flux error (see \citealt{medvedev2022}).

\begin{figure*}
    \begin{center}
    	\vspace{6mm}
    	\includegraphics[width=17cm]{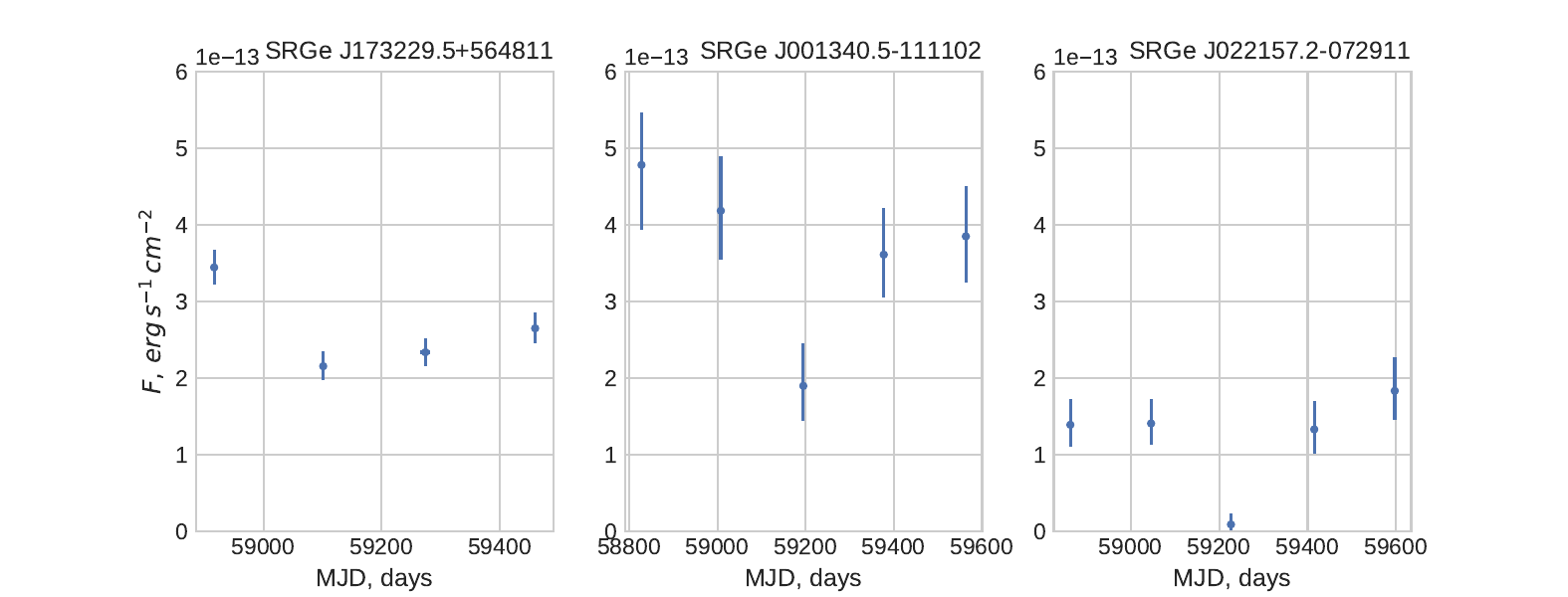}
    	\caption{Examples of eROSITA X-ray light curves. Left panel: A quasar at high ecliptic latitude ($>70^{\circ}$) and hence long exposure per visit. Middle panel: A quasar at low  ecliptic latitude ($<20^{\circ}$). Right panel: A quasar that experienced a strong drop in flux in one of the \srg\ visits.}
        \label{eRexamples}
    \end{center} 
\end{figure*}

Figure~\ref{eRexamples} shows examples of eROSITA light curves of quasars from our sample. During the \srg\ all-sky survey, any source is visited approximately every 6 months, with the duration of a single visit depending on the ecliptic latitude \citep{sunyaev2021}. Namely at low latitudes, visits typically last between one and two days, while for the highest-latitude objects in our sample this duration can reach 20 days. As can be seen from Fig.~\ref{eRexamples}, for such sources, the flux uncertainties are significantly smaller due to the longer exposure time. 

\subsection{XMM-Newton subsample}
\label{s:xmm}

\begin{figure*}
    \begin{center}
    	\vspace{6mm}
    	\includegraphics[width=17cm]{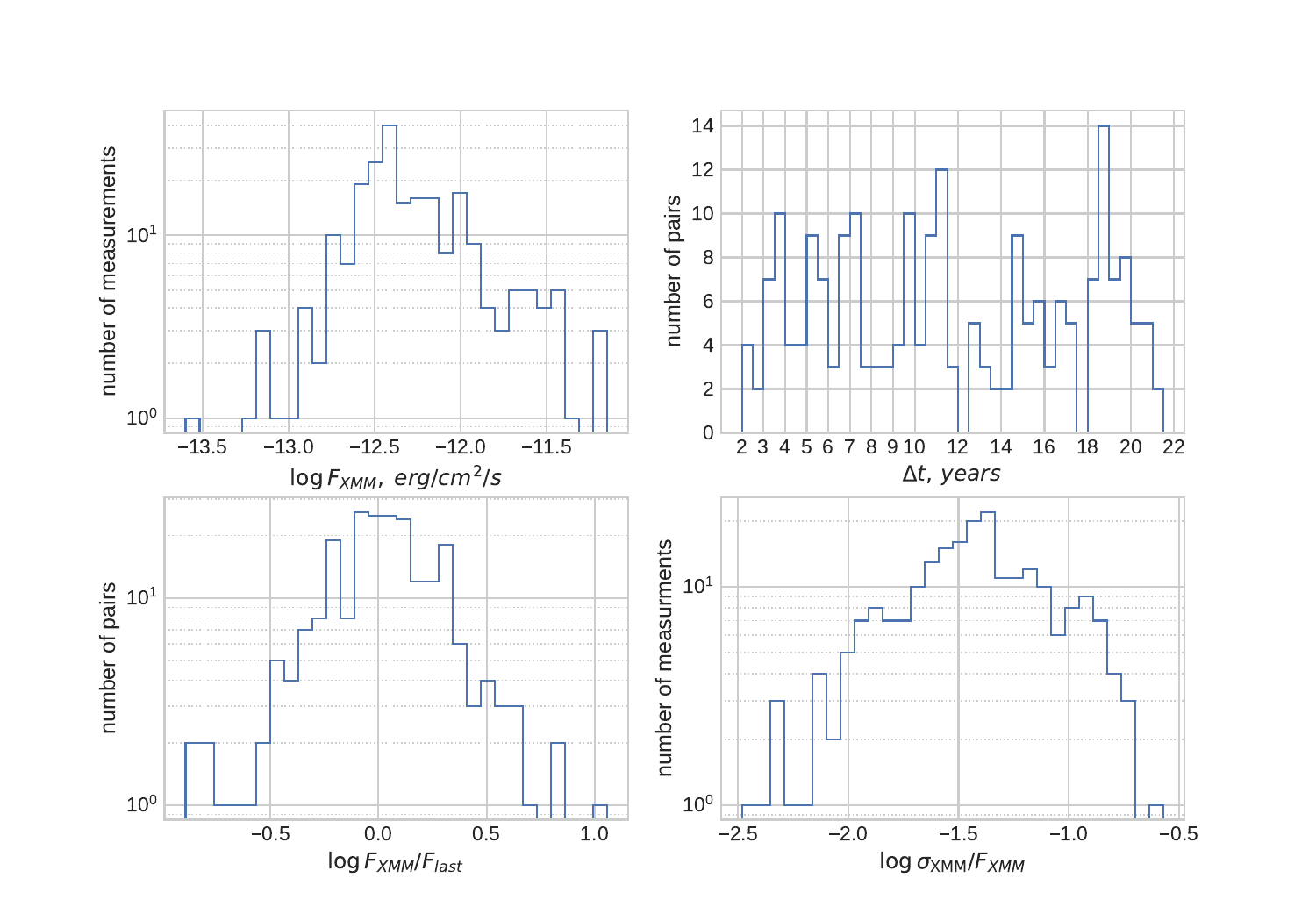}
    	\caption{Differential distributions of various X-ray properties of the \xmm\ subsample. Upper left panel: flux in the \xmm\ isolated (see the main text) observations. Upper right panel: time gap between the latest eROSITA observation and all available \xmm\ observations. Lower left panel: Ratio of the fluxes in the \xmm\ observations and in the latest eROSITA observation. Lower right panel: Relative flux measurement uncertainty in the \xmm\ observations.}
        \label{eR4XMMflux} 
    \end{center}
\end{figure*}

\begin{figure*}
    \begin{center}
    	\vspace{6mm}
    	\includegraphics[width=17cm]{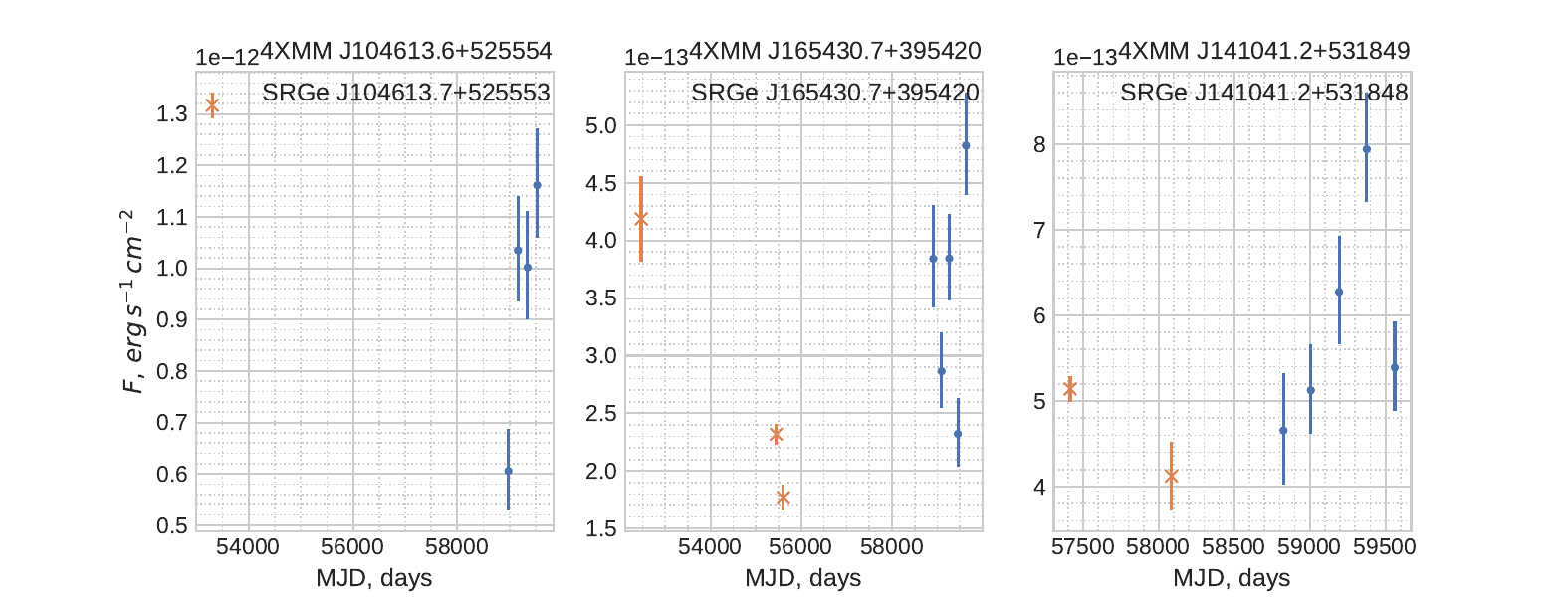}
    	\caption{Examples of long-term X-ray light curves of quasars based on \srg/eROSITA (blue dots) and \xmm\ (orange crosses) data.}
        \label{eRXMMcurves}
    \end{center}
\end{figure*}

The \srg/eROSITA data allows us to study the X-ray variability of quasars over time spans of at most 2 years. To extend our analysis to longer time scales (up to $\sim 20$ years), we cross-correlated our sample with the \xmm\ Serendipitous Source Catalogue 4XMM-DR12 \citep{webb2020xmm}, which contains X-ray flux measurements taken since the year 2000 and covers 3.1\% of the sky (for net exposure time $\ge 1$\,ks) with a typical sensitivity of a few $\times 10^{-15}$\,erg\,s$^{-1}$\,cm$^{-2}$. We conducted a search using the optical (SDSS DR14Q) positions of the quasars and the 98\% localisation regions of the \xmm\ sources, defined by radius $r_{XMM}98$. We evaluated the latter as $r_{XMM}98=1.98\times POSERR$, where $POSERR$ is provided in the 4XMM-DR12 catalogue and describes the 63\% localisation region (the factor 1.98 in the formula above corresponds to a two-dimensional Gaussian distribution). 

The cross-correlation yielded 157 matches, which constitutes 6.7\% of the \srg/eROSITA--SDSS X-ray bright quasar sample. The expected number of spurious 4XMM-DR12 counterparts is $0.3$ (it is less than 1 with a probability of 95\%). This value was found by assigning small angular offsets to the 4XMM-DR12 sources and then counting the cross-matches with our quasar sample. We excluded observations with the $SUM\_FLAG$ value greater than two, as suggested by the \xmm\ team. This resulted in 156 quasars with at least one good-quality \xmm\ flux.

Although the typical sensitivity of the 4XMM-DR12 catalogue is some two orders of magnitude better than the flux limit of our sample of X-ray bright quasars, it is in principle possible that some of our objects, due to their variability, were observed but not detected by \xmm. There is no information on such non-detections in the 4XMM-DR12 catalogue but it is available via the HILIGT tool\footnote{\url{http://xmmuls.esac.esa.int/upperlimitserver/}} \citep{konig2022hiligt}. We thus used this service to search for \xmm\ flux upper limits for our quasar sample and did not find any. Therefore, all of our quasars were sufficiently bright to be detected by \xmm\ whenever they fell into the field of view of its instruments.

We used \xmm\ fluxes in the soft (0.2--2.0\,keV) energy band, which are the sum of the EPIC fluxes listed in the 4XMM-DR12 catalogue in bands 1, 2 and 3 (0.2--0.5, 0.5--1 and 1--2\,keV, respectively). These {\it observed} fluxes had been obtained from the count rates assuming an absorbed power-law spectral model with a photon index $\Gamma= 1.7$ and a Galactic absorption column density of $\nh=3 \times 10^{20}$\,cm$^{-2}$. However, because these fluxes were determined in narrow energy ranges, they are only weakly dependent on the assumed spectral shape, e.g. the differences between the fluxes for $\Gamma=1.7$ and $\Gamma=2$ will not exceed one per cent\footnote{We checked this using the web tool \url{https://heasarc.gsfc.nasa.gov/cgi-bin/Tools/w3pimms/w3pimms.pl}.}. Hence, this difference will not exceed 1\% for the observed flux in the 0.2--2\,keV band either. 

To enable direct comparison with the eROSITA fluxes, we converted the observed 0.2--2\,keV \xmm\ fluxes to unabsorbed fluxes in the (observed) 0.3--2.3\,keV energy band assuming an absorbed power-law model with $\Gamma=2$ and $\nh=3\times 10^{20}$\,cm$^{-2}$. This corresponds to a multiplication factor of 1.15. This coefficient would change just slightly, to 1.12, if we adopted $\Gamma=1.7$. Similarly to the \srg/eROSITA data, variations in the Galactic absorption towards the different quasars in the \xmm\ sample are expected to induce $\sim 10$\% variations in flux absorption correction, which we neglect. It is worth noting again absorption correction does not affect the inferred variability characteristics of sources. 

The 4XMM-DR12 catalogue, composed of serendipitous source detections, is characterised by irregular timing structure, in contrast to the \srg\ all-sky survey data. To approximately mimic the \srg/eROSITA observation strategy, we performed exposure weighted averaging of the individual fluxes measured by \xmm\ in observations conducted within 5 days of each other. We also found the corresponding mean dates, $t_{\rm XMM}$, of such sets of observations. Hereafter, we refer to these merged observations as `isolated' \xmm\ observations, or simply as \xmm\ observations. We chose the maximum interval of 5 days because series of observations of a given target by \xmm\ usually do not span longer periods, and indeed very few consecutive \xmm\ flux measurements in our sample are separated by times between 5 and 25 days (rather than by much longer times of several years). 

There are 109 of the 156 quasars that have only one isolated \xmm\ observation, 32 have 2, 8 have 3, and 7 have 4 isolated observations. Figure\,\ref{eR4XMMflux} illustrates the main properties of the \xmm\ subsample of the \srg/eROSITA--SDSS X-ray bright quasar sample. Specifically, we show the distributions of: (i) fluxes in \xmm\ observations, (ii) time lags between the latest eROSITA visit and all \xmm\ observations of a given source, (iii) ratios of the fluxes in \xmm\ observations and in the latest eROSITA observation, and (iv) the relative uncertainties of \xmm\ fluxes. We see that the \xmm\ and eROSITA fluxes, despite the scatter caused by variability, are generally consistent with each other. The median value of the logarithm of their ratio is $0.017\pm 0.024$ (the uncertainty was estimated by a bootstrap method), i.e. consistent with zero. We also note that the vast majority of the \xmm\ fluxes are measured very precisely, to better than 10\%. Figure~\ref{eRXMMcurves} shows examples of long-term X-ray light curves of our quasars as measured by \xmm\ and \srg/eROSITA.

\subsection{Physical properties of the quasars}
\label{Physprop}

\begin{figure*}
    \begin{center}
    	\vspace{6mm}
    	\includegraphics[width=0.95\textwidth]{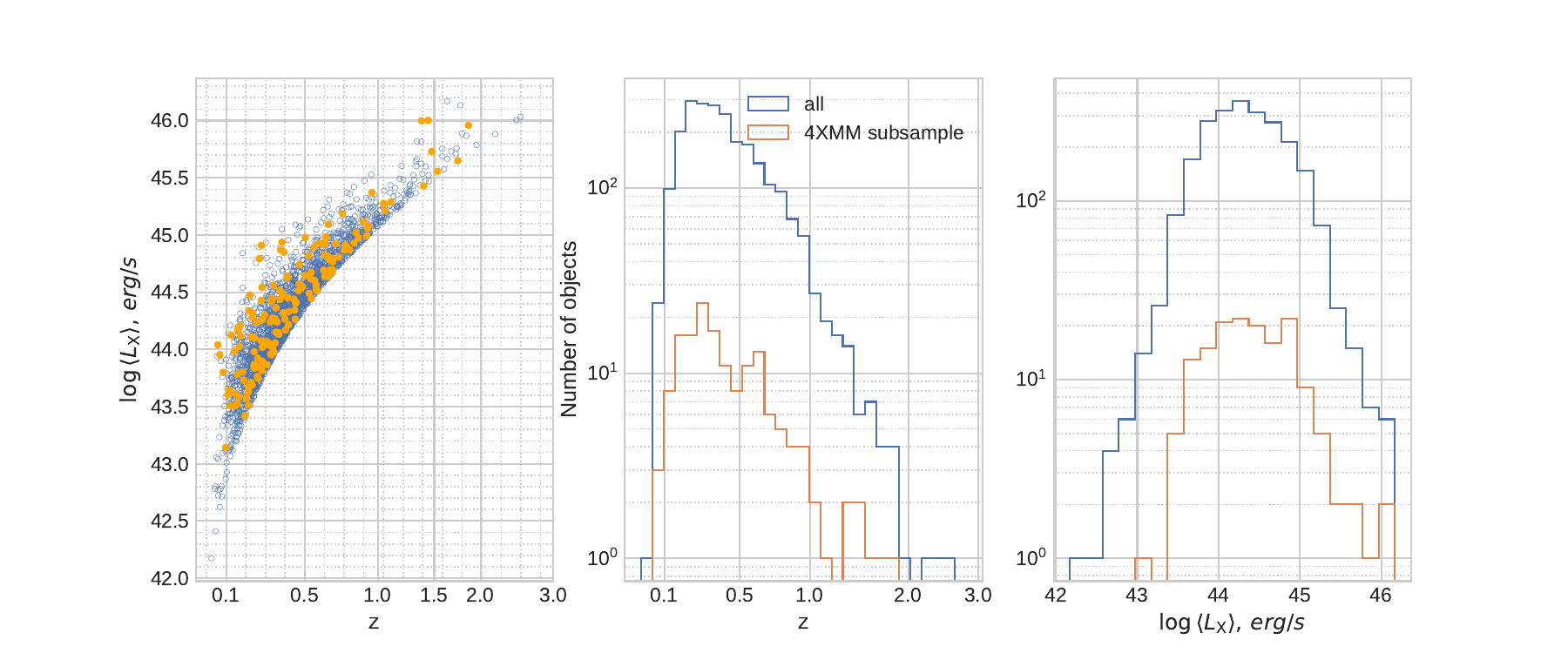}
    	\caption{Average X-ray (2--10\,keV) luminosity during the \srg/eROSITA survey vs. redshift (left), as well as the corresponding distributions of the redshifts (middle) and X-ray luminosities (right) of the studied quasars. The whole sample is shown in blue and the \xmm\ subsample in orange.}
        \label{samplezLx} 
    \end{center}
\end{figure*}

\begin{figure*}
 \begin{center}
	\vspace{6mm}
	\includegraphics[width=0.95\textwidth]{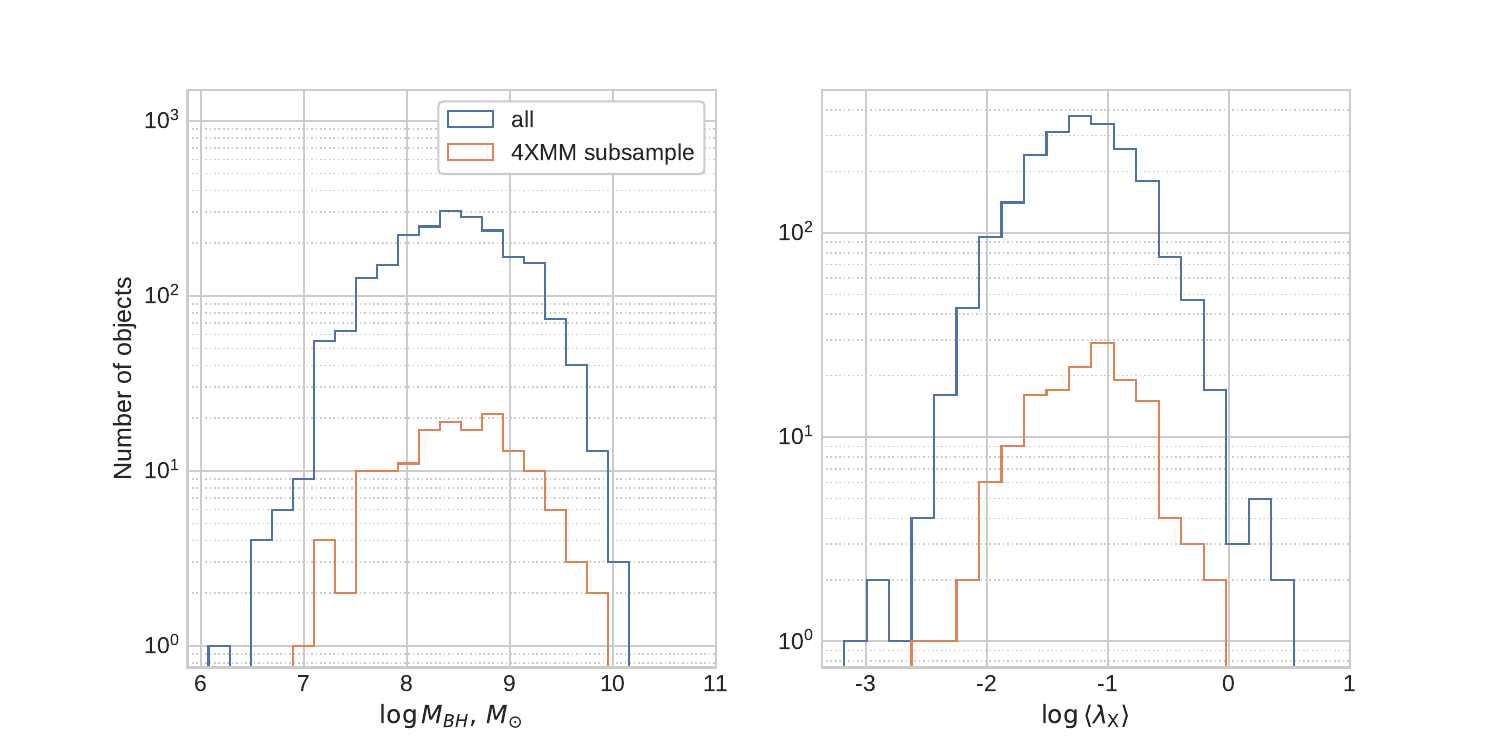}
	\caption{Distributions of the BH masses (left) and Eddington ratios (right) of the studied quasars. The whole sample is shown in blue and the \xmm\ subsample in orange.}
    \label{sampleMBHLedd}
\end{center} 
\end{figure*}

The SDSS DR14Q catalogue \citep{paris2018} provides the spectroscopic redshifts of the quasars under consideration. Ninety per cent of our objects have redshifts between 0.15 and 0.97, and the median redshift is $z_{\rm median}=0.38$. Having this information, we can determine the X-ray luminosities of the quasars. To facilitate comparison with previous studies, we define luminosities in the rest-frame 2--10\,keV energy band, $\Lx$. To this end, we use the unabsorbed fluxes in the observed 0.3--2.3\,keV band 
and calculate the $K$-correction for a power-law spectrum with $\Gamma=1.8$, which is a typical slope of (type 1) AGN spectra in the standard X-ray band (e.g. \citealt{brightman2013,trakhtenbrot2017}). 

Due to the range of redshifts in our sample, we actually examine the X-ray variability of different quasars in somewhat different rest-frame energy ranges. Namely, the observed X-ray range of 0.3--2.3\,keV corresponds to 0.4--3.2\,keV in the rest-frame at $z_{\rm median}=0.38$, whereas for 90\% of our objects the probed energies vary between 0.35--2.65 and 0.6--4.5\,keV. Therefore, our results might be affected by intrinsic dependence of X-ray variability on energy (due to spectral shape variability or variable intrinsic absorption), if there is any. Some previous studies suggested that the energy dependence is not substantial. In particular, \cite{ponti2012} found that variability in the soft (0.7--2\,keV) and hard (2--10\,keV) energy bands was tightly correlated and of similar amplitude on time scales up to 80\,ks (where is was probed) for their \xmm\ sample of AGN.

Hereafter, we denote by $\Lxs$ the luminosity of a given quasar determined from its mean eROSITA flux $\Fmeanx$. The median value of $\Lxs$ for the sample is $10^{44.3}$\,erg\,s$^{-1}$. The sample average statistical uncertainty of $\log\Lxs$ is 0.03, which directly stems from the sample average uncertainty of $\log\Fmeanx$ (since we neglect any uncertainties in the spectroscopic redshift measurements and have adopted the same $K$-correction for all the objects, see above). Figure~\ref{samplezLx} shows the luminosity--redshift diagram as well as the distributions of $z$ and $\Lxs$ for the whole \srg/eROSITA--SDSS X-ray bright quasar sample and for the \xmm\ subsample. We see that we effectively probe $\sim 2.5$\,dex in X-ray luminosity (from $\sim 10^{43}$ to $\sim 10^{45.5}$\,erg\,s$^{-1}$) and that the \xmm\ subsample has similar redshift and luminosity properties as the entire sample.

We also made use of the catalogue of spectral properties of quasars from SDSS DR14Q \citep{rakshit2020}, which, in particular, provides estimates of the BH masses, $\Mbh$. These are based on the optical continuum luminosity and line width measurements from single-epoch SDSS spectroscopy. The authors used parameters of strong emission lines such as ${\rm H}\beta\,\lambda{\rm 4861}$, ${\rm Mg\,II}\,\lambda{\rm2798}$ and ${\rm C\,IV}\,\lambda {\rm 1549}$ wherever available. It is important to note that although the formal statistical uncertainties of the resulting $\Mbh$ estimates are small, the systematic uncertainty associated with the underlying empirical relations is significant, $\sim 0.4\,$dex (e.g. \citealt{collin2006systematic, shen2013mass}), and is unknown for a given object. Only one of the 2344 quasars in our sample has no $\Mbh$ estimate in the \cite{rakshit2020} catalogue. In addition, the estimates for 108 quasars are of bad quality, according to the QUALITY\_MBH flag in that catalogue. We thus exclude these 109 quasars from those considerations below where the value of the black hole mass is required.

To characterise the regime of accretion for a given quasar, we are interested in knowing its Eddington ratio, $\ledd=\Lbol/\Le$. Here, $\Lbol$ is the bolometric luminosity of the quasar and $\Le=1.3\times10^{38}(\Mbh/\Msun)$\,erg\,s$^{-1}$ is its Eddington luminosity. 
 
Determining $\Lbol$ is an untrivial task, because it involves applying a bolometric correction for a given spectral range and can be significantly affected by variability. In particular, although we could adopt the optically-based $\Lbol$ estimates for our objects provided by \cite{rakshit2020}, we decided to refrain from that because the SDSS spectral observations were typically carried out $\sim 10$\,years before the \srg\ survey and may thus have caught a given quasar in a significantly different luminosity state compared to the period of time over which we study its X-ray variability. Instead, we estimated the bolometric luminosities directly from the eROSITA X-ray measurements as $\Lbol=\kx\Lx$, using a constant bolometric correction $\kx=10$ for the 2--10\,keV energy band, based on \cite{sazonov2012}. This allows us to define an X-ray based Eddington ratio $\lx=\kx\Lx/\Le$ and the corresponding average quantity $\lxs=\kx\Lxs/\Le$. We emphasize that the so-derived $\lx$ is a crude proxy of the true Eddington ratio, since the $\Lx/\Lbol$ ratio can actually depend on the BH mass and the Eddington ratio, as is actively discussed in the literature (e.g. \citealt{vasudevan2007,vasudevan2009,lusso2012,bongiorno2016,duras2020}). We do not try to take these dependencies into account to avoid circular reasoning.

Figure~\ref{sampleMBHLedd} shows the distributions of $\Mbh$ and $\lxs$ for the whole sample and for the \xmm\ subsample. Again, the \xmm\ subsample is similar in terms of BH masses and Eddington ratios to the rest of the sample. It is worth noting that the observed distributions of X-ray luminosities, BH masses and Eddington ratios shown in Figs.~\ref{samplezLx} and \ref{sampleMBHLedd} are, of course affected by selection effects. In particular, there is a paucity of low-luminosity quasars because we are dealing with a flux limited sample. This implies that for lower-luminosity quasars, mostly located at  lower redshifts, we are probing on average somewhat longer rest-frame time scales than for higher-luminosity quasars. This should affect our estimates of the uncertainties of the variability characteristics. However, we do not expect any significant impact of the selection effects on our conclusions regarding the variability trends with physical parameters, since our analysis is based on examining variability properties within fairly narrow  $\Mbh$, $\Lx$ and $\lx$ bins, without any cross-talk between these subsamples.

\subsubsection{Radio properties}
\label{radio}

\begin{figure*}
\begin{center}
	\vspace{6mm}
	\includegraphics[width=17cm]{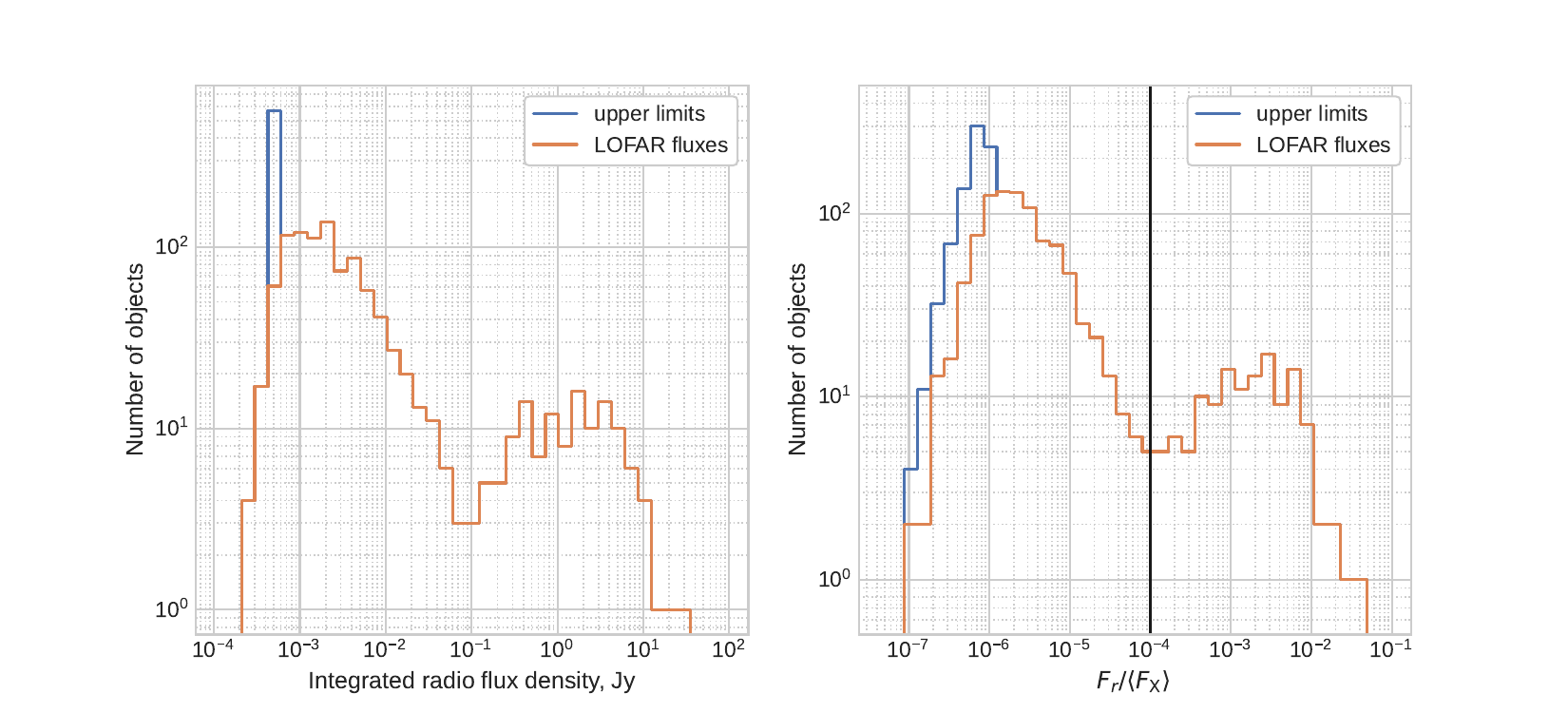}
	\caption{Left panel: Distribution of the radio flux densities of the studied quasars, as measured by LOFAR in the 120–168\,MHz band. The orange histogram shows actual measurements, while the blue column (on top of the histogram) indicates the number of non-detections with an estimated flux density upper limit of 0.45\,mJy. Right panel: The corresponding distribution of the ratios of the LOFAR to eROSITA fluxes (the upper limits are shown on top of the actual measurements). The black vertical line is drawn at $\Fr/\Fmeanx=0.0001$, which we define as a dividing line between radio-quiet and radio-loud quasars.}
    \label{sampleradio}
\end{center} 
\end{figure*}

\begin{figure*}
\begin{center}
	\vspace{6mm}
	\includegraphics[width=17cm]{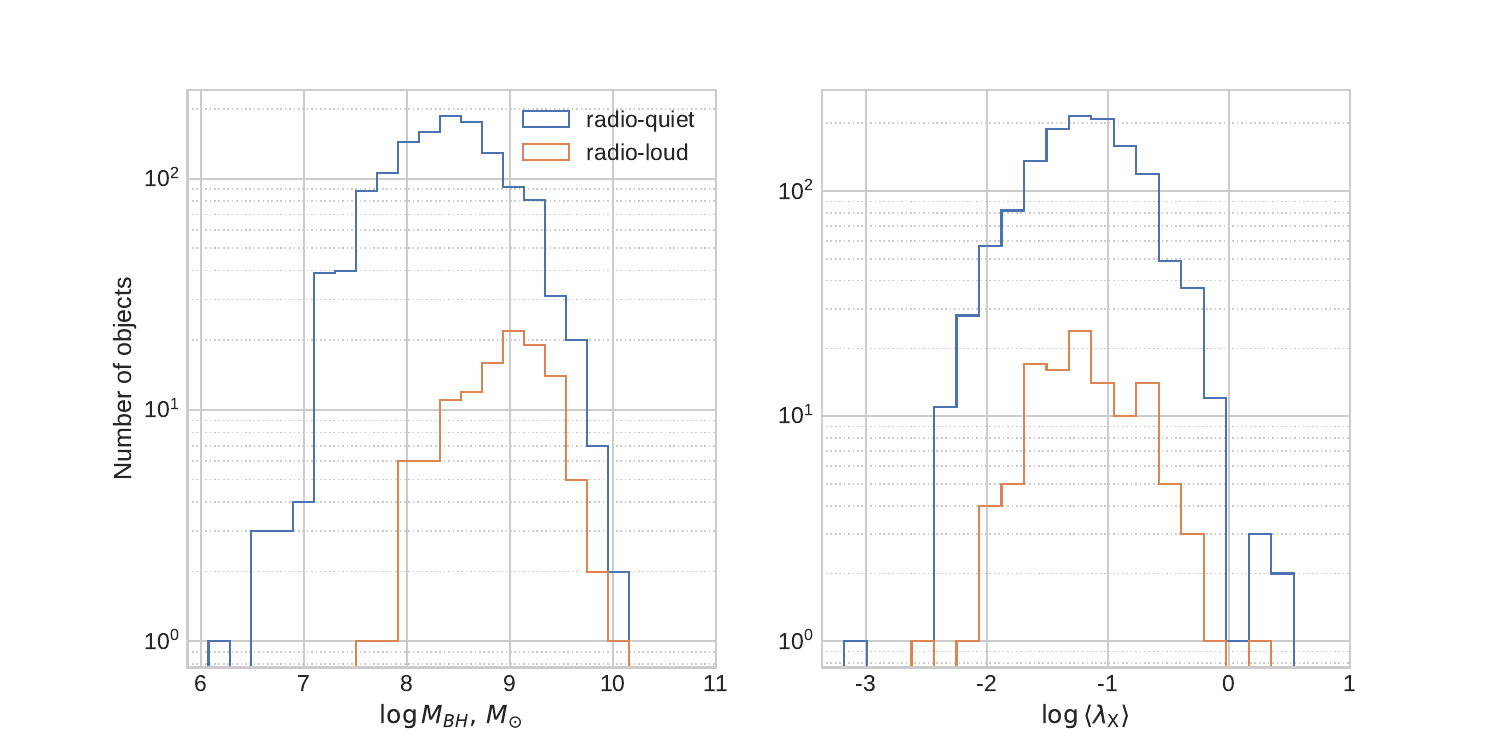}
	\caption{Distributions of the BH masses (left) and Eddington ratios (right) of the radio-quiet (blue) and radio-loud (orange) quasar samples in the LoTSS footprint.}
    \label{loudnessMBHLedd}
\end{center} 
\end{figure*}

It is also interesting to investigate if the X-ray variability of quasars depends on their radio-loudness, which in turn may be related to BH spin (e.g. \citealt{sikora2007,tchekhovskoy2010}) or to the magnetic flux threading the BH \citep{sikora2013}. To this end, we cross-correlated our sample with the LOw-Frequency ARray (LOFAR) Two-metre Sky Survey DR2 (LoTSS) \citep{shimwell2022lofar}, which has covered 27\% of the northern sky with good angular resolution (6\arcsec) and high sensitivity in the 120–168\,MHz band. The point-source completeness of the catalogue is $\sim 90$\% at an integrated (over the solid angle) flux density of 0.45\,mJy. Taking into account that some of the radio counterparts of quasars may be extended, we used a radius of 10\arcsec\ for cross-matching between the \srg/eROSITA--SDSS X-ray bright quasar sample (using the optical coordinates) and LoTSS. 

A total of 1542 quasars from the \srg/eROSITA--SDSS X-ray bright quasar sample are located in the footprint of LoTSS. Of these, 1035 have at least one LOFAR counterpart, while 13 quasars have two LOFAR counterparts. We assume that the latter are physically related pairs (e.g. double-lobed radio sources) and thus simply add up their fluxes. In what follows, we compute the radio flux from the LoTSS integrated flux density, $F_{\nu}$, as $\Fr=F_{\nu}\times 48\,{\rm MHz}$, since 48\,MHz is the width of the LoTSS frequency band. If no radio counterpart is found for a quasar located in the area covered by LoTSS, we assign an upper limit to its radio flux of $0.45\,{\rm mJy}\times 48\,{\rm MHz}$, since this is the effective sensitivity of LoTSS (see above). The expected number of spurious LOFAR counterparts is $36\pm6$. This value was estimated by assigning small angular offsets to all LoTSS sources and counting the number of cross-matches within 10\arcsec\ of our quasars. Although the number of spurious radio counterparts is fairly large (which is caused by the high density of LoTSS objects on the sky), it is nevertheless just $\sim 2$\% of the total number of counterparts and thus cannot significantly affect our results. 

Figure~\ref{sampleradio} shows the distributions of radio fluxes and radio to X-ray flux ratios for the quasars located in the LoTSS footprint. We see a bimodal shape, well known from previous studies (e.g. \citealt{sikora2007}). The minimum of the distribution between the two peaks approximately lies at $\Fr/\Fmeanx= 0.0001$, which is convenient to define as the boundary between radio-quiet and radio-loud quasars. With this definition, 128 quasars (8.3\%) turn out to be radio-loud and the remaining 1414 quasars are radio-quiet.

Figure~\ref{loudnessMBHLedd} shows the BH mass and Eddington ratio distributions for the radio-quiet and radio-loud samples of quasars located in the LoTSS footprint. We see that the radio-loud sample is dominated by heavy ($\Mbh\sim 10^9\,\Msun$) BHs, whereas the $\lxs$ distributions are similar for the radio-quiet and radio-loud samples.

\section{Variability analysis}
\label{s:analysis}

Our goal is to investigate the dependence of X-ray variability on different rest-frame timescales from $\sim$~half a year to $\sim20$\,years and on various physical parameters, namely $\Lx$, $\Mbh$, $\lx$ and radio-loudness. Since we have very few X-ray data points for each quasar but a large sample of such objects, we necessarily base our analysis on ensemble averaging.

As a measure of X-ray variability, we utilize the structure function (SF). It is based on pairs of flux measurements taken at different moments in time and is well suited for studying large samples of objects with poorly sampled light curves. The SF has often been used in astronomy, with the exact definition slightly varying from one work to another (e.g. \citealt{simonetti1985,diclemente1996,vandenberk2004, vagnetti2011}). Here we define it as follows (see e.g. \citealt{press1992,kozlowski2016,vagnetti2016}):
\begin{align}
\label{SF}
 SF^2(\Delta t) &=\left< \left[\log\frac{\Ftrue(\tobs+\dtobs)}{\Ftrue(\tobs)}\right]^2 \right>\nonumber\\
                &=\left< \left[ \ftrue(\tobs+\dtobs)-\ftrue(\tobs)\right]^2 \right>,
\end{align}
where the angle brackets indicate the ensemble average, $\Ftrue$ is the true (i.e. that would be measured in the absence of noise) X-ray flux of a given object at a given moment, $\ftrue\equiv\log\Ftrue$, and the rest-frame and observed-frame (hereafter denoted with a subscript ``obs'') time intervals are related as follows:
\begin{equation}
\label{dt}
 \Delta t=\dtobs/(1+z).
\end{equation} 
We thus characterise the variability by the squared difference between the logarithms of the fluxes, or equivalently by the square of the logarithm of the X-ray flux (i.e. luminosity) ratio, on a given rest-frame time scale. 

To ensure the mutual statistical independence of $SF^2$ estimates in different $\Delta t$ bins, we should exclude from the averaging in equation~(\ref{SF}) those pairs of flux measurements that can be algebraically expressed through other pairs of measurements for the same object that are already used in the calculation (see also \citealt{emmanoulopoulos2010}). To this end, we build each pair from the latest eROSITA observation and some previous observation by eROSITA or \xmm. Hence, for each quasar, we use all the available eROSITA and \xmm\ observations only once, expect for the latest eROSITA observation, which is used in every pair of flux measurements involved in the averaging. We have adopted the latest eROSITA observation as a reference one because this allows us to exploit the maximum available time scale for each object. 

Due to the Poisson statistics of photons, the detector actually measures a flux $\Fx$ that is different from the true flux $\Ftrue$. In logarithmic terms this corresponds to $f\equiv\log\Fx=\ftrue+\delta f$, where $\delta f$ is a random fluctuation. We thus need to subtract the contribution of photon statistical noise to the SF. To this end, it is reasonable to assume that flux fluctuations ($\delta f$) are uncorrelated with the corresponding fluxes ($\ftrue$) and with themselves\footnote{Note that due to the Poisson nature of fluctuations there is, of course, a correlation between the absolute value of the fluctuation ($|\delta f|$) and the flux ($f$), but fluctuations can be positive and negative.}. Then, we may compute a one-point estimate of $SF^2$ for each (i'th) pair of flux measurements, corrected for the noise (see \citealt{press1992}): 
\begin{align}
\label{wi}
\wi &=\left[f_i(t_{{\rm obs},i}+\Delta t_{{\rm obs},i})-f_i(t_{{\rm obs},i})\right]^2-\sigma_i^2(t_{{\rm obs},i})\nonumber\\
&-\sigma_i^2(t_{{\rm obs},i}+\Delta t_{{\rm obs},i}),
\end{align}
where $\sigma_i^2(t_{{\rm obs},i})$ and $\sigma_i^2(t_{{\rm obs},i}+\Delta t_{{\rm obs},i})$ are calculated from the corresponding measured X-ray flux uncertainties as follows:
\begin{equation}
\label{sigma}
\sigma=\frac{\sigmax}{\Fx\ln{10}}.
\end{equation}

As noted in Section~\ref{s:data}, the vast majority of sources in the \srg-SDSS X-ray bright quasar sample have at least 20 net source counts, so that their flux errors should be Gaussian to a first approximation. In order to check the error propagation expressed by equation~(\ref{sigma}), we performed a simple simulation. We assumed a Poisson distribution of counts and computed the standard deviation of $\log\Fx$. We found that the true standard deviation of $\log\Fx$ is 4\% (2\%) higher than the estimate given by equation~(\ref{sigma}) for 20 (30) counts. We then repeated this simulation for a normal distribution based on individual flux measurements and the corresponding errors from our data and obtained deviations of 7.5\% and 4\%, respectively. This minor inaccuracy in the rms calculation is unlikely to significantly affect the results of this work.

We then compute the SF for a given ensemble of flux measurement pairs as follows:
\begin{equation}
\label{SFfin}
 SF^2=\frac{1}{\npair}\sum_{i=1}^{\npair}\whi,
\end{equation}
where $\npair$ is the number of independent measurement pairs and $\whi$ is defined as follows. If a particular quasar has just one pair of measurements that falls into the $\Delta t$ bin under consideration, then $\whi=\wi$. If several pairs of measurements for the same quasar fall into the same time difference bin, we calculate $\whi$ as the mean $\wi$ of those pairs. Thereby we intend to avoid attributing too much weight to some quasars compared to others. However, we have also tried to treat all pairs of measurements of the same quasar as independent estimates of $\wi$, which led to negligible changes in the results. 

\section{Results}
\label{s:results}

In what follows, we study the dependence of $SF^2$ on the quasar rest-frame time difference $\Delta t$ for different subsamples of \srg/eROSITA--SDSS X-ray bright quasars. We first construct these dependencies in binned form. To this end, we divide a given subsample into $\Delta t$ bins of 0.3\,dex width in the range from $-0.9$ to 1.2\,dex. We then count the number of independent flux measurement pairs (or equivalently quasars) in each bin. If there are less than 10 pairs in some bin, we merge it with the adjacent bins until the number of pairs in the merged bin reaches 10 or the width of the merged bin reaches 1.2\,dex. If upon re-binning there still remain bins that contain less than 10 pairs, we exclude them from the subsequent analysis. This is aimed at increasing the reliability of ensemble-averaging in equation~(\ref{SFfin}) and leads to just a minor loss of data. We finally use bootstrapping to estimate the sampling distribution of the mean value of $\whi$ in each $\Delta t$ bin, i.e. the sampling distribution of $SF^2 (\Delta t)$, and evaluate the corresponding $68\%$ confidence intervals for $SF^2 (\Delta t)$. Specifically, we use 10,000 re-samples in each bin. In Appendix~\ref{appendix:A}, we provide further details of this procedure and present examples of actually measured $\whi$ distributions.

We then seek to parameterise the obtained $SF^2(\Delta t)$ dependencies, using ML estimation. The bootstrap-derived $SF^2$ sampling distribution around the actually measured $SF^2$ value in a given bin can be regarded as the probability density of the uncertainty in measuring $SF^2$ in that bin: $\rho (\delta SF^2)$. We can then assume that, once we have specified some model $SF^2_{\rm model} (\Delta t)$, the probability of measuring a value $SF^2$ in the experiment is determined by $\rho (SF^2-SF^2_{\rm model})$. We can then define the likelihood function as follows:
\begin{equation}
\label{fitlike}
\mathcal{L}(\pmb{SF^2}| \pmb{\theta},\pmb{\Delta t})=\prod_{j}p_j(SF^2_i| \pmb{\theta},\Delta t_j)=\prod_{j}\rho_{j}(SF^2_j-SF^2_{\rm model}(\pmb{\theta},\Delta t_j)),
\end{equation}
where $\pmb{SF^2}=\left\{SF^2_j\right\}$ is the data set, $\pmb{\theta}$ are the parameters of the model $SF^2_{\rm model}$, and the product is taken over all $\Delta t_j$ bins. 

\subsection{Models}
\label{s:models}

\begin{figure*}
\begin{center}
	\vspace{6mm}
	\includegraphics[width=0.9\textwidth]{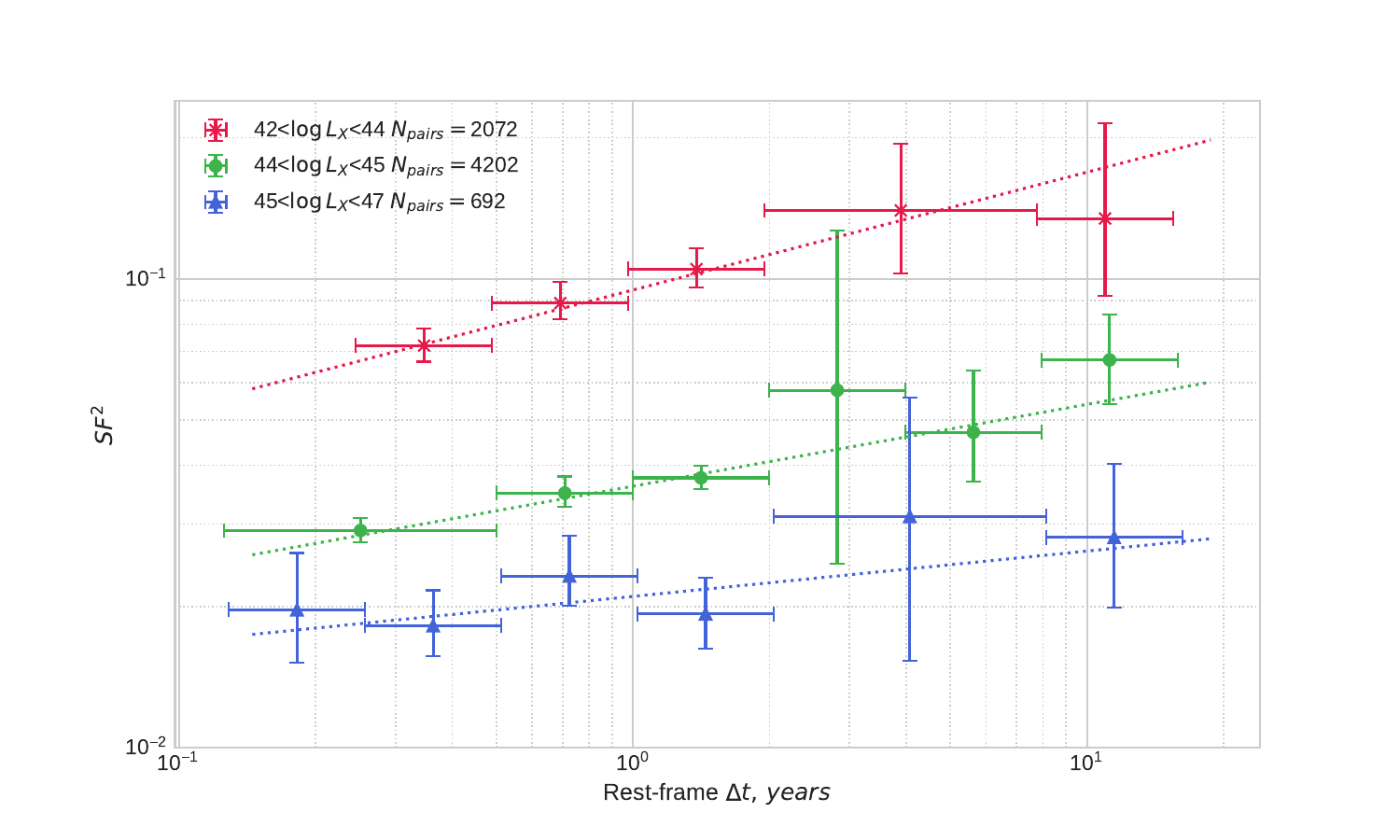}
	\caption{$SF^2$ as a function of the rest-frame time scale for different X-ray luminosity subsamples. Error bars indicate $68\%$-confidence intervals. The number of flux measurement pairs ($\npair$) used in the calculation for each subsample is quoted in the legend. The different dependencies are slightly shifted along the $\Delta t$ axis with respect to each other for better visibility. The dotted lines show fits power laws (equation~\ref{sfpl}). The best-fitting parameter values are given in Table~\ref{parametersl}.}
    \label{SF(Lx)} 
\end{center}
\end{figure*}

\begin{table*}
	\vspace{2mm}
	\centering
	\vspace{2mm}
	\begin{tabular}{c|c|c|c|c} \hline\hline
		$\log{\Lx(\rm erg\,s}^{-1})$ & median $\log{\Lx(\rm erg\,s}^{-1})$ & $N$ of $\Delta t$ bins & $A$ & $\beta$ \\ 
 \hline

$(42, 44)$ & 43.77 & 5 & $0.302\pm0.015$ & $0.25\pm0.07$ \\
$(44, 45)$ & 44.43 & 6& $0.191\pm0.004$ & $0.16\pm0.04$ \\
$(45, 47)$ & 45.18 & 6 & $0.144\pm0.005$ & $0.07\pm0.11$ \\

 \hline
	\end{tabular}
  \caption{Parameters of the best-fitting $SF^2 (\Delta t)$ functions by power laws in different X-ray luminosity bins. The parameter uncertainties are given at the $1\sigma$ confidence level.}
  \label{parametersl} 
\end{table*}

A number of previous studies of quasar variability, in particular, in the optical band \citep{macleod2010,zu2013,kozlowski2016}, used `decorrelation' models to describe the $SF^2 (\Delta t)$ dependence. A particular case of this class of models corresponds to the damped random walk (DRW) process. On the other hand, numerous studies of Seyfert galaxies (e.g. \citealt{uttley2002,uttley2005a,mchardy2004,vaughan2005,gonzalez2012}) usually found a bending power law, to satisfactorily describe the observed X-ray PSDs of AGN. In this model, the PSD follows a power law (PSD$(\nu)\propto\nu^{-\alpha}$) with a slope of $\alpha\approx 1$ at low frequencies and a steeper power law ($\alpha\approx 2$) at high frequencies. The characteristic time scale varies between a few minutes and a few years from one Seyfert galaxy to another and appears to be proportional to BH mass {(see \citealt{paolillo2023} and references therein). A bending power-law model might also be suitable for describing the $SF^2(\Delta t)$ dependence of quasar X-ray variability. However, the relationship between PSD and SF is not-trivial (e.g. \citealt{emmanoulopoulos2010}).

Both the decorrelation model and the bending power-law model are characterized by three or more parameters. Upon multiple trials, we realized that the statistical quality of our data is not sufficient for constraining them simultaneously. Moreover, we found that $SF^2 (\Delta t)$ for different studied susamples can be satisfactorily described by a simple power law:
\begin{equation}
\label{sfpl}
SF^2(\Delta t)=A^2\,\Delta t^{\beta}.
\end{equation}
Here, $\Delta t$ is measured in years so that $A$ characterises the amplitude of the variability on a time scale of 1\,year. 
In more specific terms, the aforementioned models involving a characteristic time scale do not provide a statistically significant improvement in the fit quality (e.g. based on the Akaike information criterion). Therefore, in what follows we discuss $SF^2 (\Delta t)$ dependencies in terms of the power-law model given 
by equation~(\ref{sfpl}). 

\begin{figure*}
\begin{center}
	\vspace{6mm}
	\includegraphics[width=0.9\textwidth]{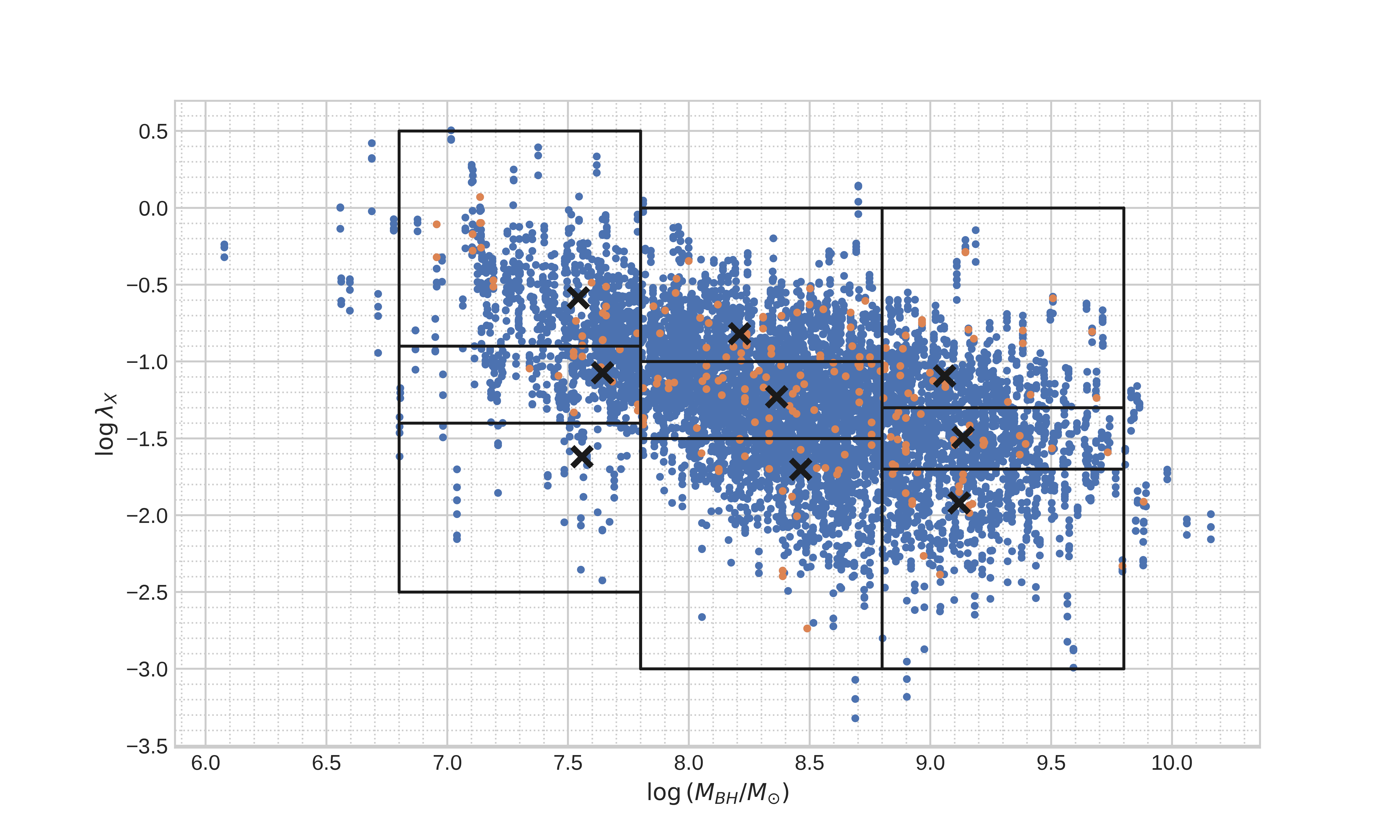}
	\caption{Scatter plot of $\Mbh$ vs. $\lx$ for the independent flux measurement pairs used in our analysis. Orange dots denote \srg/eROSITA--\xmm\ pairs. The black lines divide the ($\Mbh$, $\lx$) space into 9 regions that have been used for constructing the $SF(\Delta t)$ dependencies shown in Fig.~\ref{SF(Ledd,MBH)}. The black crosses indicate the median $\Mbh$ and $\lx$ values for each bin.}
    \label{LeddvsMBH}
\end{center} 
\end{figure*}

\subsection{Luminosity dependence}
\label{s:sflum}

We first investigate the dependence of variability on X-ray luminosity. To this end, we divide our sample into three subsamples: $\prange[\Lx]{42}{44}$, $\prange[\Lx]{44}{45}$, and $\prange[\Lx]{45}{47}$\,erg\,s$^{-1}$. Hereafter, by $\Lx$ for a flux measurement pair we mean the arithmetic mean of the X-ray luminosities of the two measurements involved. 

The results are presented in Fig.~\ref{SF(Lx)}. We see that X-ray variability tends to increase with increasing time scale and, for a given time scale, with decreasing luminosity. The binned $SF^2(\Delta t)$ dependencies can be satisfactorily described by a power law, whose best-fit parameters are given in Table~\ref{parametersl}. 

\subsection{Dependence on the BH mass and Eddington ratio}
\label{BHmARdep}

We next investigate the dependence of X-ray variability on BH mass and Eddington ratio. To this end, we divide our sample into nine subsamples: $\prange[\lx]{-2.5}{-1.4}$, $\prange[\lx]{-1.4}{-0.9}$ and $\prange[\lx]{-0.9}{0.5}$ for $\prange[\Mbh]{6.8}{7.8}\,\Msun$; $\prange[\lx]{-3}{-1.5}$, $\prange[\lx]{-1.5}{-1}$ and $\prange[\lx]{-1}{0}$ for $\prange[\Mbh]{7.8}{8.8}\,\Msun$; and $\prange[\lx]{-3}{-1.7}$, $\prange[\lx]{-1.7}{-1.3}$ and $\prange[\lx]{-1.3}{0}$ for $\prange[\Mbh]{8.8}{9.8}\,\Msun$. These subsamples have been selected so (see Fig.~\ref{LeddvsMBH}) that: (i) they contain sufficiently many data points, and in particular \srg/eROSITA--\xmm\ pairs (corresponding to long time scales), to enable an adequate statistical analysis, (ii) they are well separated from each other in terms of their median $\Mbh$ and $\lx$ values (as indicated by black crosses in Fig.~\ref{LeddvsMBH}), and (iii) there are several  $\lx$ bins per nearly the same median $\Mbh$, so that it is possible to set apart the dependencies of X-ray variability on $\lx$ and $\Mbh$. Note that a small fraction of the quasars, in particular those with very low ($\Mbh\sim 10^{6.5}\,\Msun$) or very high ($\sim 10^{10}\,\Msun$) BH masses, have remained outside of the adopted ($\Mbh$, $\lx$) regions. Also note that $\lx$ for a given flux measurement pair is computed from the mean of the corresponding X-ray luminosities. 

\begin{figure*}
\begin{center}
	\vspace{6mm}
	\includegraphics[width=0.9\textwidth]{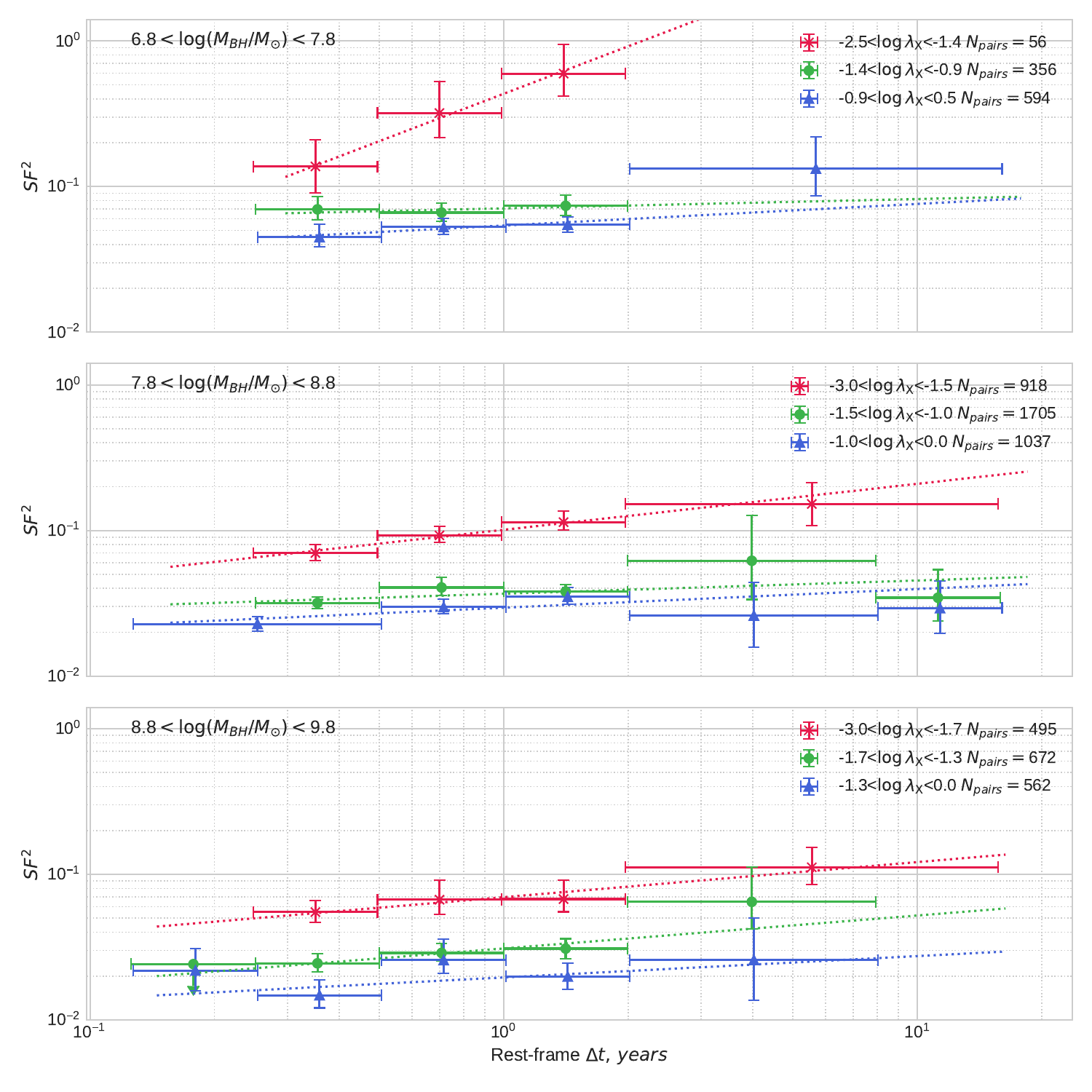}
	\caption{$SF^2$ as a function of the rest-frame time scale for different ($\Mbh$, $\lx$) subsamples. Downward arrows indicate $84\%$-confidence upper limits. The dotted lines show fits by power laws. The best-fitting parameter values are given in Table~\ref{parametersm}.}
    \label{SF(Ledd,MBH)}
\end{center} 
\end{figure*}

\begin{table*}
	
	\vspace{2mm}
	\centering
	
	\vspace{2mm}
	\begin{tabular}{c|c|c|c|c|c|c} \hline\hline
 
		$\log\Mbh(\Msun)$ & $\log{\lx}$ & median $\log\Mbh(\Msun)$  & median $\log{\lx}$ & $N$ of $\Delta t$ bins & $A$ & $\beta$\\ 
        \hline   

$(6.8, 7.8)$ & $(-2.5, -1.4)$ & 7.56 & -1.62 & 3 & $0.7\pm0.2$ & $1.1\pm 0.2$ \\
$(6.8, 7.8)$ & $(-1.4, -0.9)$ & 7.64 & -1.07 & 3 & $0.268\pm0.015$ & $0.06\pm0.19$ \\
$(6.8, 7.8)$ & $(-0.9, 0.5)$ & 7.54 & -0.58 & 4 & $0.232\pm0.010$ & $0.16\pm0.14$ \\
$(7.8, 8.8)$ & $(-3.0, -1.5)$ & 8.46 & -1.70 & 4 & $0.317\pm0.013$ & $0.32\pm0.10$ \\
$(7.8, 8.8)$ & $(-1.5, -1.0)$ & 8.36 & -1.23 & 5 & $0.190\pm0.006$ & $0.09\pm0.07$ \\
$(7.8, 8.8)$ & $(-1.0, 0.0)$ & 8.21 & -0.82 & 5 & $0.171\pm0.006$ & $0.14\pm0.06$ \\
$(8.8, 9.8)$ & $(-3.0, -1.7)$ & 9.12 & -1.92 & 4 & $0.264\pm0.016$ & $0.24\pm0.14$ \\
$(8.8, 9.8)$ & $(-1.7, -1.3)$ & 9.13 & -1.49 & 5 & $0.173\pm0.008$ & $0.22\pm0.13$ \\
$(8.8, 9.8)$ & $(-1.3, 0.0)$ & 9.06 & -1.09 & 5 & $0.142\pm0.009$ & $0.15\pm0.15$ \\

\hline
\end{tabular}
    \caption{Parameters of the best-fitting $SF^2 (\Delta t)$ functions by power laws in different ($\Mbh$, $\lx$) bins.}
    \label{parametersm} 
\end{table*}

The results are presented in Fig.~\ref{SF(Ledd,MBH)} and Table~\ref{parametersm}. There is a clear trend of increasing variability with increasing time scale for low $\lx$. This trend also persists for medium and high $\lx$, but becomes less pronounced (the power-law slope $\beta\lesssim 0.3$) and is in fact just marginally detected. Also, for a given time scale, variability tends to increase with decreasing $\Mbh$ and $\lx$ and becomes especially strong for the lowest $\lx$. 

To further verify that these results are not significantly affected by our particular ($\Mbh$, $\lx$) binning, we calculated the first-order partial correlation coefficient (see e.g. \citealt{vagnetti2011}) between individual variability estimates $\whi$ and the corresponding $\lx$ values, taking the dependence on $\Mbh$ into account, for our whole sample (excluding the few quasars without $\Mbh$ estimates). We obtained a value of $-0.223$ for this coefficient and the corresponding $p$-value of $\sim 10^{-88}$. Similarly, we studied the dependence of $\whi$ on $\Mbh$, taking into account the influence of the accretion rate. We obtained the corresponding partial correlation coefficient value of $-0.187$ with the $p$-value of\footnote{ These very low p-values should be regarded as rough estimates since they have been determined using a standard, Student’s t-distribution based formula, which is accurate only in the case of normally distributed variables, which is not true in our case.} $\sim 10^{-61}$. We can thus conclude that the anti-correlation between X-ray variability and accretion rate as well as the anti-correlation between X-ray variability and $\Mbh$ are robustly established.

\subsection{Dependence on radio-loudness}
\label{s:sfradio}

\begin{figure*}
\begin{center}
	\vspace{6mm}
	\includegraphics[width=0.9\textwidth]{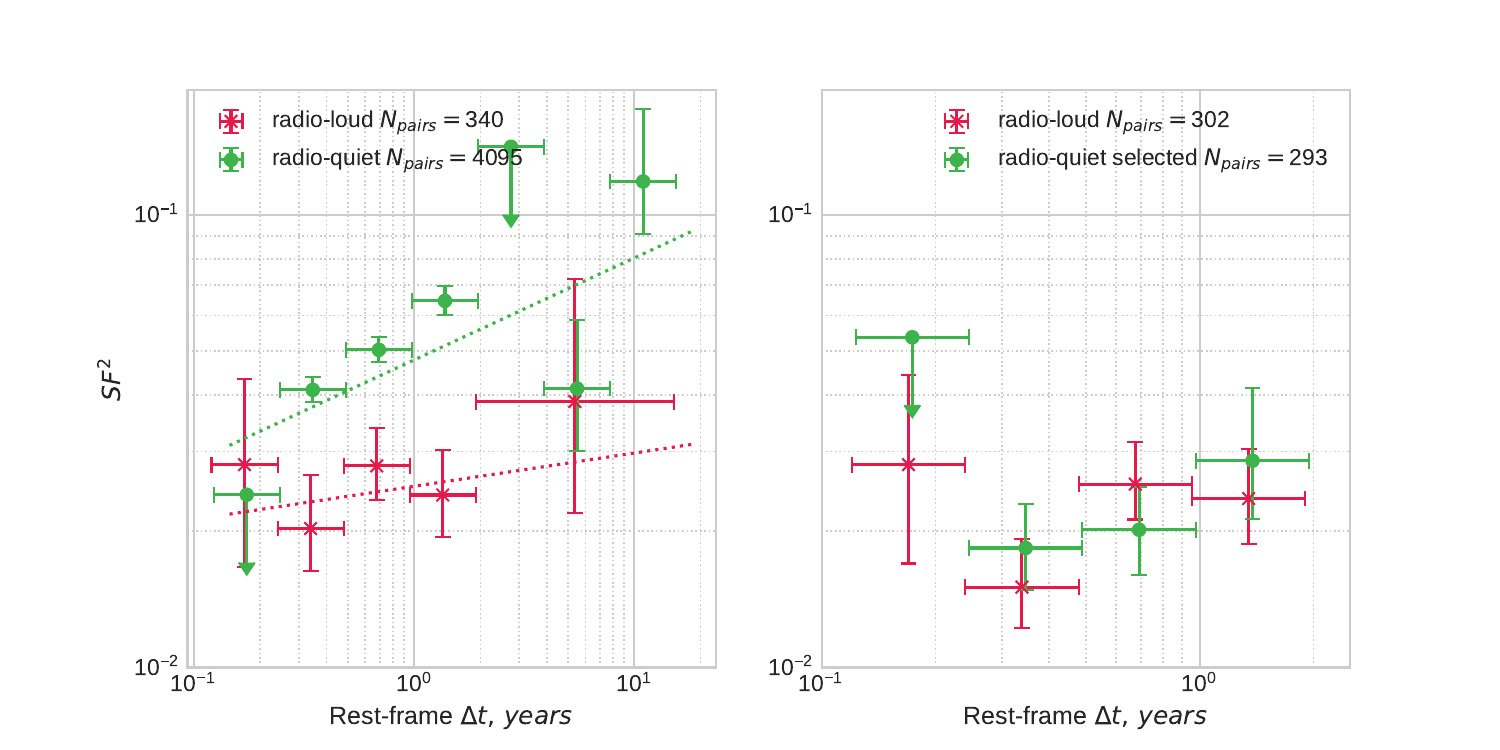}
	\caption{Left panel: $SF^2$ as a function of the rest-frame time scale for the radio-loud and radio-quiet quasar samples. The dotted lines show fits by power laws. The best-fitting parameter values are given in Table~\ref{parametersl}. Right panel: The same $SF^2 (t)$ dependence for the radio-loud sample vs. the $SF^2 (t)$ dependence for a subsample of radio-quiet quasars that is similar in terms of $\Mbh$ and $\lx$ distributions to the radio-loud sample. Note that in the right panel, we only use quasars that have good $\Mbh$ estimates for both the radio-loud and radio-quiet subsamples to check that their $\Mbh$ and $\lx$ distributions are similar. This is the reason for the slightly different number of pairs for the radio-loud samples in the left and right panels.}
    \label{SF(loudness)}
\end{center} 
\end{figure*}

\begin{table*}
	
	\vspace{2mm}
	\centering
	
	\vspace{2mm}
	\begin{tabular}{c|c|c|c} \hline\hline
		  Radio & $N$ of $\Delta t$ bins & $A$ & $\beta$ \\ 
 \hline

loud & 5 & $0.160\pm0.007$ & $0.11\pm0.12$ \\
quiet & 7 & $0.217\pm0.005$ & $0.23\pm0.03$ \\

        \hline
    \hline
	\end{tabular}
  \caption{Parameters of the best-fitting $SF^2 (\Delta t)$ functions by power laws for the radio-loud and radio-quiet quasar samples.
}
  \label{parametersr} 
\end{table*}

We finally investigate the dependence of X-ray variability on radio-loudness, based on the samples of radio-quiet and radio-loud quasars defined in Section~\ref{radio}. As shown in the left panel of Fig.~\ref{SF(loudness)}, the variability amplitude increases with increasing time scale for radio-quiet quasars. The $SF^2(\Delta t)$ dependence for radio-loud quasars is measured only on times scales shorter $\sim 1.5$\,years, where it is consistent with a constant value.

The radio-loud subsample is dominated by heavy BHs (see Fig.~\ref{loudnessMBHLedd}), which may be the cause of their apparently higher variability. To examine whether radio-quiet quasars are intrinsically more variable than radio-loud ones, or this is just another manifestation of the anti-correlation between BH mass and the amplitude of X-ray variability, we constructed a subsample of radio-quiet quasars that have $\Mbh$ and $\lx$ distributions similar to those of the radio-loud sample. We achieved this by collecting as many radio-quiet quasars in each ($\Mbh$, $\lx$) bin ($0.2\,$dex in $\Mbh$ and $0.25\,$dex in $\lx$) as there are radio-loud ones. Here, we use only quasars that have reliable $\Mbh$ estimates. We show the resulting $SF^2(\Delta t)$ dependence for the selected radio-quiet subsample along with the $SF^2(\Delta t)$ dependence for the radio-loud sample in the right panel of Fig.~\ref{SF(loudness)}. They are not significantly different from each other. This implies that $SF^2$ does not depend on radio-loudness once $\Mbh$ and $\lx$ selection effects are removed.

In addition, we have computed the second-order partial correlation coefficient (see \citealt{vagnetti2011}) between $\whi$ and radio-loudness of quasars, which accounts for the $\Mbh$ and $\lx$ dependencies, and found it to be equal to $0.027$ with the corresponding p-value of $0.06$. This further indicates that there is no significant intrinsic dependence of X-ray variability on radio-loudness.

\section{Discussion}
\label{s:discuss}

We now discuss our results in the broad context of AGN variability developed in previous studies. As we already discussed in Sections~\ref{s:intro} and \ref{s:results}, one common approach to studying X-ray variability of AGN, applicable when well-sampled light curves are available, consists of measuring the PSD as a function of frequency. In the past, it has mostly been applied to nearby and relatively low-luminosity AGN such as Seyfert galaxies. These studies have revealed, in particular, that higher luminosity sources tend to be less variable on short time scales compared to lower luminosity ones but this difference becomes barely noticeable at the longest time scales probed (e.g. \citealt{markowitz2004}). 

Typically, it is difficult to obtain high-quality X-ray light curves, and hence PSDs, for more distant and luminous AGN. Therefore, X-ray variability studies of quasars have usually focused on examining the dependence of integral characteristics of X-ray variability, such as fractional root mean square (rms) variability (or, equivalently, normalized excess variance) on SMBH physical parameters, and mostly employed an ensemble-averaging approach. In one of the first detailed studies of X-ray variability of quasars, \cite{papadakis2008}, using a sample of 66 objects at $z\sim 1$ observed by \xmm\ in the Lockman Hole, confirmed the trend (albeit with a large scatter) of decreasing variability amplitude with increasing X-ray luminosity on an observed time scale of $\sim 2$\,months.

\cite{yang2016} and \cite{paolillo2017} performed similar studies on longer observed time scales of $\sim 15$\,years, using samples of distant quasars from the \chandra\ Deep Field-South Survey, and solidified the existence of anti-correlation between X-ray luminosity and variability. 
In our study, based on SF, we have reaffirmed the anti-correlation of X-ray variability with quasar luminosity on rest-frame time scales from $\sim 0.2$ to $\sim 20$\,years.

It is not the first time that AGN X-ray variability has been studied in terms of SF. In particular, \cite{vagnetti2011} and \cite{vagnetti2016} previously used \xmm\ data to investigate the dependence of SF on luminosity and BH mass on rest-frame time scales from 0.1\,day to 4\,years, whereas \cite{middei2017} studied the behaviour of SF on time scales up to 20\,years. These authors, similar to our work, approximated $SF(\Delta t)$ by a simple power-law model and obtained slopes in the range $\sim 0$--0.2 for different AGN subsamples, which are in good agreement with the slopes $\sim 0$--0.4 for $SF^2(\Delta t)$ that we infer for our different subsamples (except for the subsample with the lowest BH mass and lowest Eddington ratio, where $\beta=1.1\pm 0.4$)\footnote{The slopes of $SF^2(\Delta t)$ and $SF(\Delta t)$ are related by a factor of 2.}.

The shape of the $SF(\Delta t)$ function is expected to reflect (see e.g. \citealt{macleod2010,zu2013,kozlowski2016}) that of the underlying PSD$(\nu)$ function, although the relationship between these two properties is non-trivial (e.g. \citealt{emmanoulopoulos2010}). Specifically, if the slope of PSD$(\nu)$ changes gradually from $\alpha_1$ to $\alpha_2$ (PSD$(\nu)\propto\nu^{-\alpha}$) around some frequency $\nub$, then $SF(\Delta t)$ is also expected to bend at $\Delta t\sim 1/\nub$. 
Our study has not revealed statistically significant evidence for a change of the SF slope across the probed range of time scales. 

\subsection{Key role of the Eddington ratio}
\label{s:eddratio}

From a fundamental point of view, variability properties are expected to be determined by the combination of BH mass and Eddington ratio, as well as possibly BH spin. The interplay between BH mass and Eddington ratio in shaping the long-term X-ray variability of quasars has remained poorly understood so far, largely due the lack of sufficiently large samples of objects covering broad ranges in BH mass and luminosity, and with sufficient sampling on long ($\sim 1$--10\,year) time scales.

For nearby Seyfert galaxies and for time scales shorter than 80\,ks, an anti-correlation between variability and BH mass has been established (e.g. \citealt{oneill2005,ponti2012}). However, the dependence of variability on the accretion rate remained uncertain. \cite{papadakis2004}, \cite{gonzalez2012} and \cite{lanzuisi2014} did not find significant correlation between these quantities; \cite{oneill2005} reported an anti-correlation; whereas \cite{ponti2012} and \cite{paolillo2017} concluded that the accretion rate likely affects, in some (model dependent) way, both the break frequency and amplitude of the PSD. The \srg/eROSITA--SDSS X-ray bright quasar sample allowed us to systematically address this problem. 

Arguably, the most interesting finding of this study is that X-ray variability is substantially stronger at low Eddington ratios, $\lx\lesssim 10^{-2}$, compared to higher accretion rates, at least over the $\sim 0.2$--20\,year rest-frame time scales probed by this study. This behaviour pertains to SMBHs of various mass. The trend of decreasing variability amplitude with increasing accretion rate appears to persist, but becomes less pronounced, also when we compare moderate- and high-accretion rate objects ($\lx\sim 10^{-1.25}$ vs. $\lx\sim 10^{-0.75}$). There is thus an indication that once the Eddington ratio drops below a few per cent (with a significant uncertainty in this value due to the fact that $\lx$ is a crude proxy of the true Eddington ratio), the X-ray emission becomes substantially more variable. 
Considering the X-ray variability amplitude at a given time scale and a given Eddington ratio as a function of BH mass, we find that lighter BHs tend to be more variable on relatively short time scales (less than $\sim 2$\,years) but we cannot draw a conclusion about the dependence on longer time scales with this sample.

In Appendix~\ref{appendix:B}, we provide additional information on the 10 quasars with the lowest $\lx$ in our sample. Specifically, we present their X-ray light curves (see Fig.~\ref{lrlmcurves}) and discuss their optical spectral properties based on the literature. This information can be useful for follow-up studies of these potentially interesting objects.

\subsection{Comparison with optical variability of quasars}
\label{s:optical}

Recently, a study similar to ours, but on the {\it optical} variability of quasars, has been conducted by \cite{arevalo2023}, who also similarly used the catalogue of spectral properties of quasars from SDSS DR14Q \citep{rakshit2020}. Dividing their sample into subsamples by BH mass, \cite{arevalo2023} found the Eddington ratio and the variability amplitude to be anti-correlated, confirming previous indications \citep{macleod2010,sanchez2018,li2018}. They also found that at a given time scale, the amplitude of optical variability decreases with BH mass, with this trend being stronger on short rest-frame time scales of 30--150\,days than on longer scales of $\sim 300$\,days. 

These new results on optical variability are similar to our findings on X-ray variability, which suggests that the temporal behaviours of optical and X-ray emission are largely driven by the same physical processes near the SMBH. It is important to note, however, that the results of both studies are not directly comparable to each other, since the variability analysis in \cite{arevalo2023} is restricted to times scales shorter than 1\,year, while our work is mostly focused on longer time scales. 


\section{Summary}
\label{s:summary}

We have studied the medium- and long-term (rest-frame time scales between a few months and $\sim 20$\,years) X-ray variability of a large, uniform sample of X-ray bright quasars from the SDSS DR14Q catalogue, based on the data of the \srg/eROSITA all-sky survey complemented for $\sim 7$\% of the sample by archival data from the 4XMM-DR12 catalogue. The results of this work can be summarized as follows.

\begin{itemize}

\item We reaffirm the previously known anti-correlation of the X-ray variability amplitude with X-ray luminosity.

\item We find that more massive BHs tend to be less variable, for given Eddington ratio and time scale, than less massive ones. 

\item We find the X-ray variability amplitude to increase with decreasing Eddington ratio, and there is a clear indication that variability becomes particularly strong at very low accretion rates of $\lx\lesssim$\,a few per cent, regardless of BH mass. Quantitatively, in this regime, the X-ray variability amplitude (i.e. the typical ratio of a pair of flux measurements) is a factor of $\sim 1.8$ for the most massive BHs ($10^{8.8}<\Mbh<10^{9.8}\,\Msun$) and a factor of $\sim 4$ for the least massive ones ($10^{6.8}<\Mbh<10^{7.8}\,\Msun$) on a time scale of 1\,year. For comparison, at near-Eddington accretion rates ($\lx\lesssim 1$), the corresponding variability amplitudes are $\sim 1.4$ and $\sim 1.7$, respectively. 

\item We confirm the conclusion of previous AGN studies that the X-ray variability amplitude (expressed in terms of $SF^2$) increases with increasing time scale. The $SF^2(\Delta t)$ dependencies that we have obtained for different subsamples can be satisfactorily described by power laws, with the slopes $\sim 0$--0.4 (except for the subsample with the lowest BH mass and lowest Eddington ratio, where it is equal to $1.1\pm 0.4$). Although our data are also consistent with a DRW model and with a bending power-law model, we cannot place reliable constraints on the characteristic time. 

\end{itemize}

\section*{Acknowledgements}
SP, SS, MG and PM acknowledge support by grant no. 21-12-00343 from the Russian Science Foundation. AAS was partly supported by the project number 0033-2019-0005 of the Russian Ministry of Science and Higher Education. The authors thank the referee for helpful comments.

\section*{Data Availability}
The X-ray data analysed in this article were used by permission of the Russian \srg/eROSITA consortium. The data will become publicly available as part of the corresponding \srg/eROSITA data release along with the appropriate calibration information. 


\medskip

\bibliographystyle{mnras}
\bibliography{eRvariability}

\appendix
\section{Distribution of one-point estimates of the structure function}
\label{appendix:A}

\begin{figure*}

\begin{center}
	\vspace{6mm}
	\includegraphics[width=15.5cm]{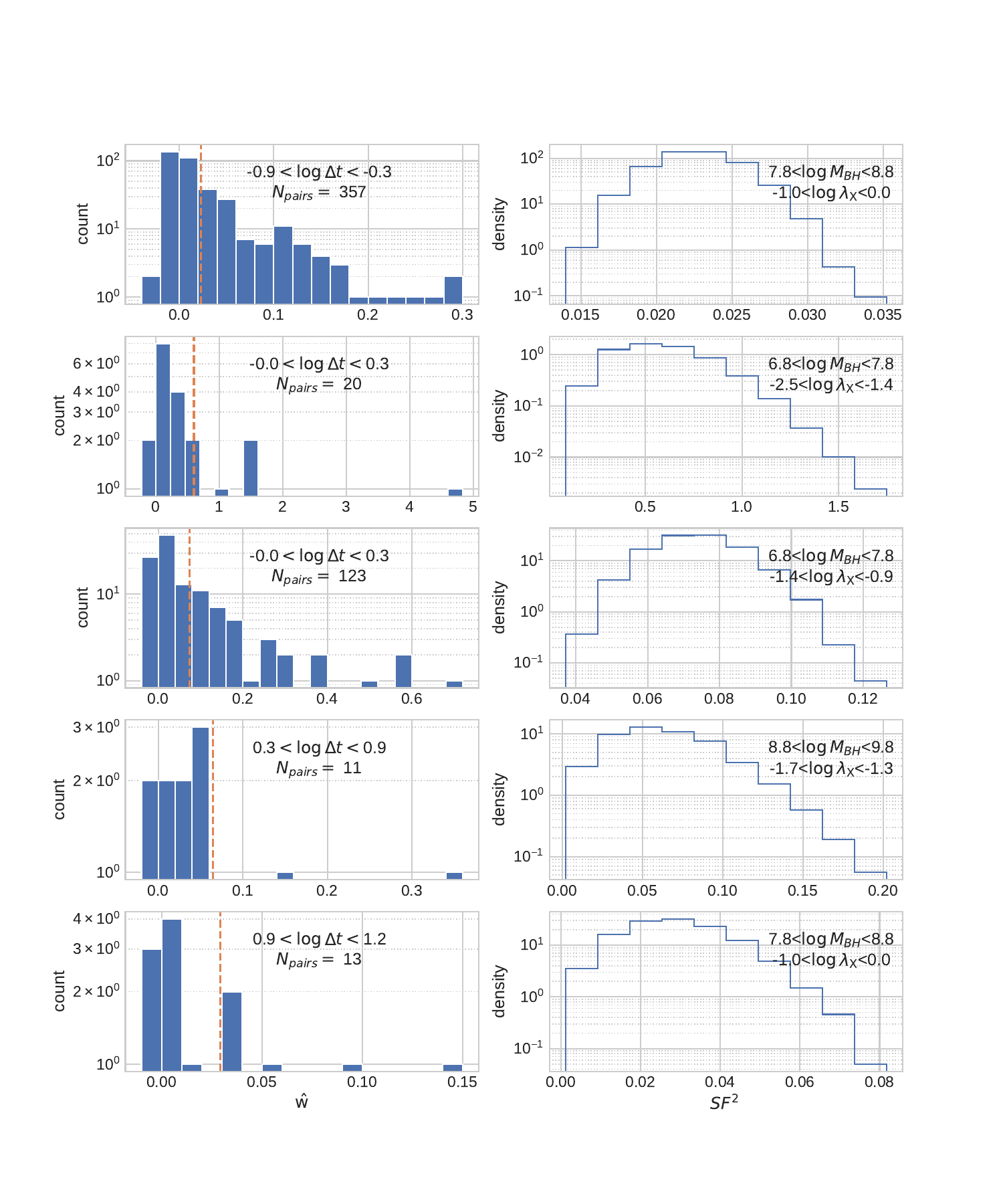}
	\caption{Left panels: the $\whi$ distributions for some of the ($\Mbh$, $\lx$, $\Delta t$) subsamples used for obtaining the $SF^2(\Delta t)$ dependencies shown in Fig.~\ref{SF(Ledd,MBH)}. The dashed orange line marks the mean $\whi$ value (i.e. the estimated $SF^2$ value) of a given distribution. Right panels: the $SF^2$ probability distributions found by bootstrapping over the $\whi$ distributions in the corresponding left panels. The legends provide information on the ($\Mbh$, $\lx$, $\Delta t$) subsamples including their size ($\npair$).}
    \label{whidistrs}
\end{center} 
\end{figure*}

In this study, we determine the structure function by ensemble-averaging one-point estimates, $\whi$, using equation~(\ref{SFfin}). Figure~\ref{whidistrs} (left panels) shows examples of actually measured $\whi$ distributions, which correspond to different ($\Mbh$, $\lx$) subsamples and different $\Delta t$ intervals that we used to obtain the $SF^2(\Delta t)$ dependencies shown in Fig.~\ref{SF(Ledd,MBH)}. 

First, we see that there is a negative part ($\whi<0$) in these distributions. This is unsurprising, since the X-ray flux uncertainties are not negligible even for the bright quasar sample studied in this work, and their estimated contribution to $\whi$ may outweigh the intrinsic flux variation in a one-point estimate (see equation~\ref{wi}). However, the median and average values of such distributions are always positive, provided the ensemble is not very small (we require at least 10 one-point estimates). Second, we see that the $\whi$ distributions are asymmetric and skewed towards large $\whi$ (i.e. towards strong flux variations).

In the right panels of Fig.~\ref{whidistrs}, we show the corresponding $SF^2$ probability distributions, derived by bootstrapping over the measured $\whi$ distributions. These $SF^2$ distributions are much more symmetric than the $\whi$ distributions. We use these simulated $SF^2$ probability distributions to estimate the 68\% confidence regions for the binned $SF^2(\Delta t)$ dependencies and to determine the likelihood function defined in equation~(\ref{fitlike}). To implement this last step, we need to convert the simulated (discrete) $SF^2$ distribution into a smooth probability distribution function $\rho(\delta SF^2)$. To this end, we use kernel density estimation based on a Gaussian kernel of some optimal width, via Scott's rule \cite{scott2015}.

\section{Variability of low-Eddington ratio AGN}
\label{appendix:B}

\begin{table*}
	
	\vspace{2mm}
	\centering
	
	\vspace{2mm}
	\begin{tabular}{c|c|c|c|c|c|c|c|c} \hline\hline
		{SRGe name}&  SDSS name & Redshift & $\log{\Lxs}$ & $\log{\Mbh}$ & $\log{\lxs}$ & $F_{\rm max}/F_{\rm min}$ & ${\rm FWHM}({\rm H}\alpha)$ & ${\rm FWHM}({\rm H}\beta)$\\ 
        &   & &  & $(\Msun)$ &  & & (1000\,km\,s$^{-1}$) & (1000\,km\,$s^{-1}$)\\
        \hline
J020151.6+012902 & J020151.65+012902.5 & 0.155 &   43.27$\pm$0.03 &   8.73$\pm$0.12 & -2.57$\pm$0.13 & 5.1$\pm$1.4 &   7.8$\pm$0.8 &  9.6$\pm$1.4 \\
J025231.2+034113 & J025231.19+034112.7 & 0.267 &   43.88$\pm$0.03 &   9.18$\pm$0.09 & -2.42$\pm$0.09 & 4.7$\pm$1.3 &   9.1$\pm$0.7 & 10.0$\pm$1.0 \\
J075403.5+481429 & J075403.60+481428.0 & 0.276 &   43.88$\pm$0.04 &   9.59$\pm$0.09 & -2.83$\pm$0.10 & 2.6$\pm$0.8 &  12.5$\pm$1.4 & 13.7$\pm$1.5 \\
J113029.3+634621 & J113029.48+634620.4 & 0.073 &   42.62$\pm$0.04 &   8.69$\pm$0.14 & -3.18$\pm$0.14 & 3.2$\pm$0.9 &      10$\pm$3 & 12.0$\pm$1.9 \\
J134617.6+622047 & J134617.54+622045.5 & 0.116 &   43.14$\pm$0.03 &   8.60$\pm$0.03 & -2.58$\pm$0.04 & 2.7$\pm$0.6 & 5.69$\pm$0.08 &  7.6$\pm$0.2 \\
J150752.7+133845 & J150752.66+133844.5 & 0.322 &   44.01$\pm$0.03 & 9.566$\pm$0.017 & -2.68$\pm$0.04 & 4.2$\pm$1.5 &  13.0$\pm$0.5 & 13.5$\pm$0.3 \\
J155053.3+052115 & J155053.16+052112.1 & 0.110 &   43.11$\pm$0.03 &   8.90$\pm$0.06 & -2.90$\pm$0.06 &   60$\pm$33 &   6.0$\pm$0.4 &  8.5$\pm$0.6 \\
J162752.2+541912 & J162752.18+541912.5 & 0.316 & 44.067$\pm$0.018 &   9.44$\pm$0.12 & -2.48$\pm$0.12 & 6.9$\pm$0.8 &      14$\pm$3 &     16$\pm$2 \\
J214611.6-070448 & J214611.58-070449.2 & 0.125 &   43.44$\pm$0.03 &   8.75$\pm$0.02 & -2.42$\pm$0.04 & 3.1$\pm$0.8 & 6.48$\pm$0.15 &  8.0$\pm$0.2 \\
J233254.4+151305 & J233254.46+151305.5 & 0.215 &   43.71$\pm$0.04 & 9.039$\pm$0.019 & -2.44$\pm$0.04 &     7$\pm$3 &   8.9$\pm$0.3 & 10.1$\pm$0.2 \\
        \hline
		
	\end{tabular}
  \caption{Information about the 10 AGN with the lowest Eddington ratios ($\lxs<0.004$) in the \srg/eROSITA-SDSS X-ray bright quasar sample.}
  \label{lowratelowmass} 
\end{table*}

\begin{figure*}
\begin{center}
	\vspace{6mm}
	\includegraphics[width=15.5cm]{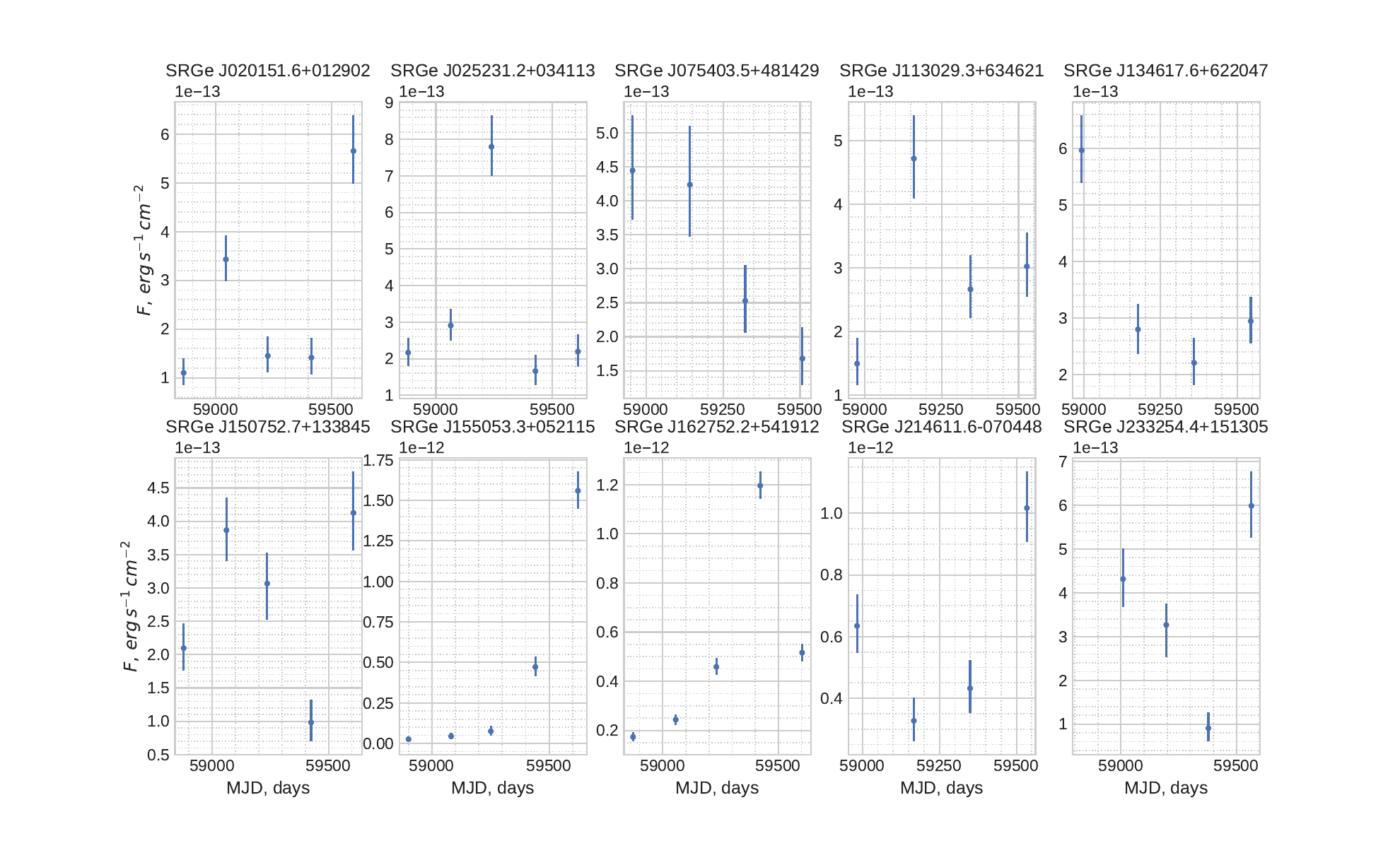}
	\caption{X-ray light curves of the 10 AGN with the lowest $\lx$, listed in Table~\ref{lowratelowmass}. Only eROSITA data are shown, because none of these objects have \xmm\ measurements.}
    \label{lrlmcurves}
\end{center} 
\end{figure*}

As was shown in Section~\ref{BHmARdep}, quasars with low Eddington ratios demonstrate on average particularly strong X-ray variability. It is worth to examine the physical properties of such objects in more detail. There are 10 quasars in our sample that have $\lxs<0.004$. Table~\ref{lowratelowmass} provides some key information about these AGN. We also show their X-ray light curves in Fig.~\ref{lrlmcurves}. 

Since all of these objects are relatively nearby ($z<0.32$) and have relatively low X-ray luminosities ($\Lxs\lesssim 10^{44}$\,erg\,s$^{-1}$), it is more appropriate to refer to them as AGN or Seyfert galaxies rather than quasars. A salient feature of these objects is that the Balmer emission lines in their optical (SDSS) spectra are very broad (see the last column in Table~\ref{lowratelowmass}, based on \citealt{rakshit2020}). This is certainly related to the fact that the broad-line region (BLR) becomes smaller with decreasing luminosity (e.g. \citealt{kaspi2005}) and the gas velocity dispersion accordingly increases by virtue of the viral theorem. However, this may be just part of the explanation, since the profiles of the Balmer lines of these objects are complex and some of them have extreme widths up to ${\rm FWHM}\sim 1.6\times 10^4$\,km\,s$^{-1}$. 

In fact, the complexity of emission line profiles in some of these AGN has attracted attention before and given rise to various interpretations. In particular,  SRGe\,J075403.5+481429 = SDSS\,J075403.60+481428.0 and SRGe\,J134617.6+622047 = SDSS\,J134617.54+622045.5 were suggested to be candidate binary SMBHs \citep{eracleous2012} based on the apparent velocity shift of the peak of the emission lines from the objects' systemic redshifts. Similarly, SRGe\,J155053.3+052115 = SDSS\,J155053.16+052112.1 was considered to be a candidate binary SMBH by \cite{2013ApJ...775...49S} based on observed changes in its optical spectrum over time, but \citet{2019MNRAS.482.3288G} later concluded that this spectral variability was  more likely caused by intrinsic changes in the BLR of a single SMBH. Another object, SRGe\,J233254.4+151305 = SDSS\,J233254.46+151305.5, was suggested to be a recoiling SMBH, i.e. the product of the merger of two SMBHs, because its active nucleus appears to be physically shifted from the center of the host galaxy by $\sim 6$\,kpc \citep{2021ApJ...913..102W}. 

Whatever the right explanation of the extreme width and profile complexity of the emission lines in these objects may be, it is plausible that the BH masses determined from standard scaling relations and quoted in Table~\ref{lowratelowmass} are overestimated for some of them, so that the corresponding Eddington ratios may need to be corrected upwards. Nevertheless, this subsample may contain a significant fraction of genuine low-$\lx$ objects.

\end{document}